\documentclass[a4paper,11pt]{article}
\usepackage{jcappub} % for details on the use of the package, please see the JINST-author-manual
\usepackage{lineno}
\usepackage{bigints}        %make integral sign bigger
\usepackage{xcolor}         %to make the text coloured
\arxivnumber{2310.19968} % Only if you have one

\title{Constraining primordial black hole masses through \boldmath{$f(R)$} gravity scalarons in Big Bang Nucleosynthesis}

\author[a]{Abhijit Talukdar,}
\author[a]{Sanjeev Kalita,}
\author[a]{Nirmali Das}
\affiliation[a]{Department of Physics, Gauhati University, Gopinath Bordoloi Nagar, 781014, India}
\author[b]{and Nandita Lahkar}
\affiliation[b]{Department of Physics, Guwahati College, Jyoti Nagar Road, Bamunimaidan, 781021, India}

\emailAdd{abhijittalukdar@gauhati.ac.in}

\abstract{Big Bang Nucleosynthesis (BBN) is a strong probe for constraining new physics including gravitation. $f(R)$ gravity theory is an interesting alternative to general relativity which introduces additional degrees of freedom known as scalarons. In this work we demonstrate the existence of black hole solutions in $f(R)$ gravity and develop a relation between scalaron mass and black hole mass. We have used observed bound on the freezeout temperature to constrain scalaron mass range by modifying the cosmic expansion rate at the BBN epoch. The mass range of primordial black holes (PBHs) which are astrophysical dark matter candidates is deduced. The range of scalaron mass which does not spoil the BBN era is found to be $10^{-16}-10^4 \text{ eV}$ { for both relativistic and non-relativistic scalarons}. {The window $10^{-16}-10^{-14}$ eV of scalaron mass obtained from solar system constraint on PPN parameter is compatible with the BBN bound derived in this work}. The PBH mass range is obtained as $10^6-10^{-14}\text{ }M_{\odot}$. Scalarons constrained by BBN are also eligible to accommodate axion like dark matter particles. The problem of ultra-light PBHs ($M \le 10^{-24} \text{ }M_\odot$) not constrained by the present study of BBN is still open. Estimation of deuterium (D) fraction and relative D+$^3$He abundance in the $f(R)$ gravity scenario shows that the BBN history mimics that of general relativity. While the PBH mass range is eligible for non-baryonic dark matter, the BBN bounded scalarons provide with an independent strong field test of $f(R)$ gravity. {The PBH mass range obtained in the study is discussed in relation to future astronomical measurements.}}


\begin{document}
\maketitle
\flushbottom
%%%%%%%%%%%%%%%%%%%%%%%%%%%%%%%%%%%%%
%%%%%%%%%%%%%%%%%%%%%%%%%%%%%%%%%%%%%
%%%%%%%%%%%%%%%%%%%%%%%%%%%%%%%%%%%%%
\section{Introduction}\label{sec1}

Primordial black holes (PBHs) were formed out of large random fluctuations in density of the primordial universe \citep{PhysRevLett.26.1344, 1974MNRAS.168..399C}. Large amplitude fluctuations {responsible for formation of PBHs can also produce cosmological gravitational waves (GWs).} Future GW observatories such as Laser Interferometer Space Antenna (LISA) are eligible to test {this idea of PBH formation} \citep{doi:10.1073/pnas.2211215119}. As the black hole formation is a classical process, it would only form if its Schwarzschild radius ($R_\text{s} = 2GM/c^2$) is greater than the Planck length {($R_\text{Pl}\approx \sqrt{{\hbar G}/{c^3}})$}. A prototype relationship {between mass of the PBHs and the time of their formation after the birth of the universe} can be obtained by equating the cosmological mass density, ${1}/{G t^2}$ with density required to form a black hole, ${3M}/{4 \pi {R_\text{s}}^3}$. {Therefore, the PBH mass as a function of cosmic time ($t$) is expressed as,

\begin{equation}
	M_\text{PBH}=m_\text{Pl}\frac{t}{t_\text{Pl}}
\end{equation}
where, $m_\text{Pl}=\sqrt{\hbar c/G}$ is the Planck mass and $t_\text{Pl}=\sqrt{\hbar G/c^5}$ is the Planck time. 
}These PBHs have masses $10^{-8}$ kg (formed at the Planck time, $t_\text{Pl} \approx 10^{-43}$ s) upwards \citep{1974MNRAS.168..399C}.

 { PBHs emit Hawking radiation due to quantum processes near the horizon, they loose masses in the process and get depleted of its surface gravity \citep{HAWKING1974}}. It leads to further increase in its rate of emission. A black hole of mass $M$ would have a finite lifetime which varies with mass as $\tau \varpropto M^3$. A PBH with mass smaller than $10^{12}$ kg  {($10^{-18}\text{ } M_\odot$)} completely evaporates by { emitting Hawking radiation within} the present cosmic time. It must have mass above $10^{12}$ kg  to have observable astrophysical effects in the present universe. The horizon of the PBHs may be of the order of the particle horizon ($c a(t) \int_0^t dt' a(t')^{-1}$,  $a(t)$ being the cosmic scale factor) at the epoch of formation. Therefore, PBHs carry a wide range of masses. { If we consider PBHs forming at the end of cosmic inflation, they carry masses of the order of $10^{-3}$ kg {($10^{-33} \text{ }M_\odot$)}}. 1$\text{ }M_{\odot}$ PBHs can form at the end of Quantum Chromodynamic (QCD) phase transition ($10^ {-6}-10^{ - 5}$ s after the Big Bang). A PBH can be as massive as $10^6\text{ }M_{\odot}$ if it forms just at the onset of Big Bang Nucleosynthesis (BBN), $t = 1$ s. { The mass of the PBHs formed depends on the mechanism that triggers their formation process \citep{Volonteri2021TheOO}}. { Thus the expected mass range of PBHs is $10^{-33}$--$10^6\text{ }M_{\odot}$. In this work we narrow down masses of the PBHs which can serve as non-baryonic dark matter and hence must survive till present epoch. Therefore, our target mass range for PBHs is $10^{-18}$--$10^6\text{ }M_{\odot}$}.  { There is possibility of growth of PBHs through accretion of relativistic particles in the early radiation era of the universe. Application of Novikov-Zeldovich accretion mechanism showed that a PBH remains in its initial mass if the size of its event horizon is less than the particle horizon. However, when the event horizon becomes as large as the particle horizon it grows catastrophically as the accretion proceeds \citep{1974MNRAS.168..399C}. Result of this study showed that PBH remains close to the initial mass near $10^{-8}$ kg (formed at the Planck time, $t_\text{Pl} \approx 10^{-43}$ s).}
 
 %There is a possibility of growth of a PBH through accretion of a relativistic particles in the radiation era. \cite{1974MNRAS.168..399C} showed that a PBH remains in its initial mass if its horizon length is smaller than the particle horizon. When it is of the order of particle horizon the black hole mass catastrophically grows with cosmic expansion as it accretes. These authors demonstrated there is no substantial growth due to accretion and a PBH remains in its original mass $10^{-8}$ kg upward.

Influence of PBHs on pre-galactic cosmological processes such as reionisation of matter and the overall thermal history of the universe were discussed earlier by \cite{1981MNRAS.194..639C}. Low mass PBHs are often discussed in the context of Hawking radiation. Although Hawking radiation of PBHs is not observationally confirmed it carries tremendous potential for explaining some puzzles such as Galactic and extragalactic $\gamma$-ray background \citep{1976ApJ...206....1P, 1976ApJ...206....8C, 1996ApJ...459..487W, Lehoucq:2009ge}, antimatter in cosmic rays \citep{Kiraly1981, 1991ApJ...371..447M} and some classes of $\gamma$-ray bursts \citep{1996MNRAS.283..626B}. This is an open problem in astronomy. Massive PBHs ($M=10^3 - 10^6\text{ }M_{\odot})$ are believed to act as seeds of supermassive black holes observed in present day galactic nuclei \citep{Rubin:2001yw, Carr:2018rid}. Existence of very massive central black holes in the high redshift quasars (e.g. $1.5 \times 10^9 \text{ }M_{\odot}$ black hole in J1007+2115 with $z >7.5$ \citep{2020ApJ...897L..14Y} and $2.2 \times 10^9 \text{ }M_{\odot}$ black hole in J1205 – 0000 with $z =6.7$ \citep{2019ApJ...880...77O}) requires massive seed black holes with mass $M \approx 10^4-10^6\text{ } M_{\odot}$ \citep{Feng:2020kxv}. It is challenging to generate these seeds through the remnants of Population III stars \citep{Inayoshi:2019fun}. Direct collapse of low metallicity primordial gas clouds is a possible channel to generate massive seed black holes \citep{Bromm:2002hb, Begelman:2006db, 2010ApJ...716.1397F}. PBHs therefore, can be a serious alternative to direct collapse black holes. { Poisson density fluctuation of the PBHs is eligible to generate large scale structure of the universe \citep{2003ApJ...594L..71A}. Therefore, PBHs are interesting components in understanding cosmic evolution. }

PBHs surviving the Hawking evaporation can constitute the non-baryonic dark matter component. The usual suspect for { non-baryonic} dark matter is a class of Weakly Interacting Massive Particles (WIMPs). { Recent studies indicate that lighter non-baryonic particles such as axions are consistent with observed less clumpy nature of the large scale structure \citep{Rogers:2023ezo}.} Fuzzy Cold Dark Matter is another idea which involves extraordinarily light scalar particles $(m \approx 10^{-22}$ eV) \citep{PhysRevLett.85.1158}. However, attempts to find new particles or fields as dark matter candidates have come up empty { handed} \citep{doi:10.1073/pnas.2211215119}. { The Laser Interferometer Gravitational Wave Observatory (LIGO) has discovered gravitational waves emitted by massive black hole binaries.  Black holes with mass $20-30 \text{ }M_{\odot}$ and above found by LIGO are beyond the upper limit of black hole masses resulting from core collapse supernovae. Therefore, these are of non-stellar origin and perhaps of primordial origin. It has raised the interest of PBHs as potential candidates for some or entire budget of non-baryonic dark matter \cite{Carr:2020xqk}. Massive PBHs are also of importance. For example, PBHs with mass $>10^3 \text{ }M_\odot$ can generate large scale structure and hence can act as serious alternative to dark matter. Planck mass relics left by evaporating PBHs and PBHs with mass $> 10^{12} \text{ } M_\odot$ are interesting dark matter candidates.  }

%After LIGO’s discovery of gravitational waves the idea of PBHs as dark matter candidates has seen a surge. Black holes with mass $20-30 \text{ }M_{\odot}$ found by LIGO are beyond the upper limit of black hole masses resulting from core collapse supernovae. Therefore, if the theory of stellar evolution is correct these are believed to be of primordial origin. Calculation of PBH formation rate in the QCD epoch has shown that the highest peak of PBH formation rate occurs near $1-2\text{ }M_{\odot}$. But there are peaks near $30-50\text{ }M_{\odot}$ \citep{PhysRevD.55.R5871, Carr:2021bzv}, similar to the mass range of LIGO black holes. The situation where PBHs can constitute some or entire dark matter mass budget of the universe has been recently reviewed by \cite{Carr:2020xqk}. PBH with mass, $M> 10^3\text{ } M_{\odot}$ can generate large scale structures. Planck mass relics of PBH evaporation and unusually large black holes with mass, $M>10^{12}\text{ }M_{\odot}$ are interesting dark matter candidates.

PBHs synthesised well before the onset of BBN ($t \approx1$ s) i.e. with masses, $M< 10^5 M_{\odot}$ are classified as non-baryonic as they are not subject to the BBN constraint on baryonic matter which contributes 5\% of the Einstein-de Sitter density. Thus, PBHs being non-relativistic and effectively collisionless can potentially serve as non-baryonic dark matter candidate \citep{Clesse:2016vqa}. PBHs in the mass windows $10^{-17}-10^{-16}\text{ } M_{\odot},\text{ } 10^{-13}-10^{-9} \text{ } M_{\odot} \text{ and } 1-10^3\text{ } M_{\odot}$ are currently being emphasised for contributing to the dark matter density \citep*{PhysRevD.94.083504}. PBHs of masses in the astrophysical window (sub-lunar to massive black hole masses, $10^{-18} -10^5\text{ }M_{\odot}$) are of potential interest. However, there may be further constraints on the lower bound of PBH mass. Sufficiently small PBHs with mass $M <<10^{15}$ kg can potentially harm the light element abundances produced in the BBN era through modification of the expansion rate caused by Hawking evaporation products \citep{PhysRevD.102.103512}. The main idea is the following. The black hole evaporation products act as density components in the Friedman-Lemaitre expansion law. They enhance the cosmic expansion rate and hence elevate the temperature of neutron-proton freezeout ($t \approx 1$ s in standard cosmology). It causes a shift in the freezeout era to earlier moments in cosmic history. This leaves one with large neutron to proton ratio which in turn elevates the primordial helium ($^4$He) abundance. Similarly, mesons as Hawking evaporation products can convert some of these protons into neutrons. It affects the abundance of deuterium (D) and $^4$He. Therefore, improved observational bounds on the light element abundances can be employed to constrain the lower side of the PBH mass range. 

%In addition to producing Hawking evaporation, black holes are also potential sites for altering the behaviour of gravity. Cosmological modification to general relativity (GR) is well motivated for understanding accelerated expansion of the primordial and the late universe.

{ In addition to producing Hawking radiation, black holes provide us with sites where gravity may behave differently from what it does in GR. Modification of GR on cosmological scale is well motivated for understanding the accelerated phases of expansion of the early and the late universe \citep*{Starobinsky:1980te, Capozziello:2003gx, Sotiriou:2008rp}.} Effect of modified gravity theories on BBN and constraints on them have been investigated through observed abundances of the light elements (D, $^4$He and $^7$Li) and observed bound on the shift of the freezeout temperature \citep{PhysRevD.93.043511, Anagnostopoulos2022gej, Asimakis:2021yct, Sultan:2022aoa}. These are alternatives to GR arising from fundamental theories of quantum gravity which are capable of generating early and late time accelerated cosmic expansion. Geometrical corrections to GR such as $f(R)$ gravity theories naturally demand existence of additional scalar gravitational degrees of freedom. These are known as scalarons. Scalarons as natural outcome of curvature corrections to quantum vacuum fluctuations near the Galactic Centre black hole have been used to forecast testability of modified gravity through pericentre shift of compact stellar orbits near the black hole and to investigate the effect of modified gravity on constraining spin of the black hole \citep{2020ApJ...893...31K, 2021ApJ...909..189K}. In these works the scalaron degree of freedom was shown to present a Yukawa type correction to gravitational potential near the black hole which affects the orbital shift of stars encircling the black hole in compact orbits. Mass of the scalaron has been related to ultraviolet (UV) and infrared (IR) cut off scales of vacuum fluctuations near the black hole \citep{2020ApJ...893...31K}. Possibility of constraining $f(R)$ gravity theories through measurements near the Galactic Centre black hole has been extensively studied in earlier investigations \citep{Borka:2015vqa, 2020ApJ...893...31K, PhysRevD.104.L101502, 2021ApJ...909..189K, 2022ApJ...925..126L, 2023EPJC...83..120K}. Cosmological tests of $f(R)$ gravity theories have been discussed earlier through large scale structure probes such as halo mass function \citep{PhysRevD.92.044009, PhysRevLett.117.051101}, mass-temperature relation of galaxy clusters \citep{2017A&A...598A.132H}, cluster gas mass fraction \citep{Li:2015rva} and clustering of clusters \citep{2017MNRAS.467.1569A}. 

{ BBN can also be used as a probe for testing strong field regime of gravitational theories. The particle horizon at the onset of BBN} ($t=1$ s) is around 0.004 au (of the order of solar radius, also see section \ref{sec2}). The cosmological density at this epoch is $\rho \approx 1.5 \times 10^8 \text{ kg m}^{-3}$. This gives the horizon mass as $M_\text{H}=4\pi \rho R_\text{H}^3/3 \approx 6.8 \times 10^4\text{ } M_{\odot}$. Therefore, the dimensionless gravitational potential, $\phi=G{M}_\text{H}/c^2R_\text{H}$ is nearly 0.2 which is five order magnitude larger than the one encountered at the neighbourhood of the Sun. This provides us with a very strong field environment to test gravity. \cite{1981ApJ...243....8F} studied production of $^4$He and deuterium in BBN as a test of Rosen's bi-metric theory of gravitation \citep{1973GReGr...4..435R, ROSEN1974455}. { Deviation from GR in the weak field limit is described by the Parametrised Post Newtonian (PPN) parameters} ($\gamma$, curvature per unit mass and $\beta$, non-linearity in gravity measured respectively by light deflection near a massive body and pericentre shift of the orbit of a test body). Solar system measurement showed that the PPN parameters of bi-metric gravity theory are identical to those in GR ($\gamma=1$, $\beta=1$). Therefore, in the weak field limit of the solar system the theory is unconstrained. Production of $^4$He and deuterium was found to be heavily suppressed by a wide class of models in this theory \citep{1981ApJ...243....8F}. This provided with strong constraint on the theory.

In the present study we work with $f(R)$ gravity and its scalaron degrees of freedom associated with black hole horizon. By extending the established relationship between mass of the scalaron and horizon size of the black holes \citep{2020ApJ...893...31K} we produce constraints on the PBH masses through available bound on the shift of the neutron-proton freezeout temperature. We treat the scalaron degree of freedom as an additional density component in the radiation era and investigate the shift of the freezeout epoch and estimate neutron fraction and deuterium abundance. Constraints on the PBH masses derived in this way are then compared with other independent constraints. The plan of this paper is as follows. In section \ref{sec2}, after demonstrating the existence of black hole solutions in $f(R)$ gravity we extend the relationship between mass of scalarons and mass of black holes and estimate scalaron masses for a wide range of expected PBH masses. In section \ref{sec3} we have discussed the formalism for constraining the mass range of $f(R)$ gravity scalarons and thereby PBH masses using observed bound on the shift of freezeout temperature. Effect of scalarons on neutron and deuterium fraction is also discussed. Finally, D+$^3$He abundance is estimated in presence of $f(R)$ gravity scalarons and compared with the observed bound. In section \ref{sec4} we present results and discussions.

\section{ \boldmath{$f(R)$} gravity and the primordial black holes}\label{sec2}

After the discovery of accelerated expansion of the universe \citep{SupernovaSearchTeam:1998fmf, SupernovaCosmologyProject:1998vns, 2007ApJ...659...98R} a mysterious `dark energy' has been proposed to account for a large scale cosmic repulsion. Einstein’s cosmological constant which serves as latent energy in empty space has been considered as a standard candidate of `dark energy'. This is an unclustered form of energy in the universe whose density remains constant with cosmic expansion. However, fine tuning problems associated with the cosmological constant \citep{RevModPhys.61.1, RevModPhys.75.559} have motivated emergence of dynamical dark energy candidates known as scalar fields. Cosmological scalar fields are inspired by the Brans-Dicke theory (the first of a class of scalar–tensor gravity theory where a scalar degree of freedom of gravity exists in addition to the metric field) \citep{PhysRev.124.925} and the paradigm of cosmic inflation \citep{PhysRevD.23.347, Linde:1983gd}. Evolution of scalar fields in cosmic time produces a state where the field develops negative pressure, thereby causing the universe to accelerate against gravity of the large scale matter distribution. Scalar field dark energy models have been extensively investigated in the context of late time cosmic acceleration \citep{PhysRevD.37.3406, PhysRevLett.80.1582, PhysRevLett.81.3067, PhysRevLett.82.896}. These are extra fields added to Einstein's field equations of gravity. Although GR has been tested with sufficient confidence in the regime of the solar system (see \citep{Will2001}), binary pulsars \citep{1982ApJ...253..908T, 1998ApJ...505..352S}, black hole binaries \citep{ PhysRevX.6.041015, PhysRevLett.116.061102} and the Galactic Centre black hole \citep{PhysRevLett.122.101102, 2018A&A...615L..15G,  2020A&A...636L...5G} the empirical basis for its extrapolation to the larger scales of cosmology is not yet { a} sufficiently strong { one}.

It has been suggested that cosmic acceleration including the primordial inflation and late time acceleration might be the result of deviation from GR \citep{Starobinsky:1980te, Starobinsky_2007,Capozziello:2002rd, Capozziello:2003gx, PhysRevD.70.043528,PhysRevD.75.127502}. There have also been attempts to unify the early inflation with late time cosmic acceleration under a common framework of modified gravity \citep{Nojiri2010wj, Nojiri:2013zza}. The class of modified gravity theories which is of cosmological origin is known as $f(R)$ gravity where the Ricci curvature scalar $R$ appearing in { the gravitational action} is replaced by a function of $R$. The field equations are derived in these theories from the modified Einstein-Hilbert action,
\begin{equation}\label{PBH1}
	S=\dfrac{c^4}{16 \pi G} \int d^4x \sqrt{-g} f(R) + S_m({\psi_m, g_{\mu \nu }}),
\end{equation}
where, $S_m$ is the action for matter fields and $\psi_m$ { are the matter fields universally coupled to the spacetime metric} $g_{\mu \nu}$. The function $f(R)$ corresponds to the scalar Lagrangian density { of the gravitational field}.  The choice  $f(R)=R$ gives rise to general relativistic field equations. Models with $f(R) \varpropto R^2$ are eligible to produce singularity free isotropic cosmology \citep{Starobinsky:1980te}. Late time accelerated expansion has been found to be possible with a low curvature { (large scale)} modification of type $f(R)=R-\alpha R^{-n} (\alpha, n>0)$ \citep{Capozziello:2002rd} and without introducing exotic negative pressure sources. There are astrophysical consequences of $f(R)$ theories. Weak field limit of the modified field equations has been shown to produce a Yukawa correction to Newtonian gravitational potential near a central mass. Testability of such corrections through in-plane pericentre shift of compact stellar orbits near the Galactic Centre black hole has been extensively investigated \citep{PhysRevD.85.124004, 2018IJMPD..2741009Z, 2020ApJ...893...31K, PhysRevD.97.104068, PhysRevD.104.L101502, 2022JCAP...03..007D}. $f(R)$ theories are also found to be eligible to explain rotation curves of galaxies without introducing non-baryonic dark matter \citep{2007MNRAS.375.1423C, 10.1007/978-3-319-02063-1_1}.

\subsection{Vacuum solution in \boldmath{$f(R)$} gravity}

Vacuum solutions of gravitational field equations are related to existence of black holes, either isolated (Schwarzschild) or with a cosmological background with constant curvature (Schwarzschild-de Sitter). In the metric formalism, the field equations in $f(R)$ gravity obtained from Hamilton’s principle,  $\delta S=0$  are \citep{amendola_tsujikawa_2010},
\begin{equation}\label{PBH2}
	f^\prime\left(R\right)R_{\alpha\beta}-\frac{f\left(R\right)}{2}g_{\alpha\beta}-\nabla_\alpha\nabla_\beta f^\prime\left(R\right)+\nabla^\mu\nabla_\mu f^\prime\left(R\right)g_{\alpha\beta}={8\pi Gc}^{-4}T_{\alpha\beta},
\end{equation}
where, $\nabla$  stands for covariant differentiation, $T_{\alpha\beta}$ is the matter-energy tensor and $f^\prime\left(R\right)=df(R)/dR$. Trace of equation \eqref{PBH2}, $g^{\alpha\beta}T_{\alpha\beta}$ gives,

\begin{equation}\label{PBH3}
	f^\prime\left(R\right)R-2f\left(R\right)+3\nabla^\mu\nabla_\mu f^\prime\left(R\right)=8\pi Gc^{-4}\left(\rho c^2-3p\right).
\end{equation}

Here $\rho$ and $p$ are density and pressure of the matter source. Existence of the term $\nabla^\mu\nabla_\mu f^\prime\left(R\right)$ in equation \eqref{PBH3} implies that there is a propagating scalar field $\psi=f^\prime(R)$. This is known as the `scalaron'. In vacuum $( \rho=0=p)$ the field equation \eqref{PBH2} become,

\begin{equation}\label{PBH4}
	f^\prime\left(R\right)R_{\alpha\beta}-\frac{f(R)}{2}g_{\alpha\beta}=\nabla_\alpha\nabla_\beta f^\prime\left(R\right)-\nabla^\mu\nabla_\mu f^\prime\left(R\right)g_{\alpha\beta}.
\end{equation}

The trace equation \eqref{PBH3} becomes,
\begin{equation}\label{PBH5}
	f\left(R\right)=\frac{f^\prime(R)R}{2}+\frac{3}{2}\nabla^\mu\nabla_\mu f^\prime(R).
\end{equation}

If one considers constant curvature background with $R=R_0$ equation \eqref{PBH5} gives,

\begin{equation}\label{PBH6}
	f\left(R_0\right)=\frac{f^\prime(R_0)R_0}{2}.
\end{equation}

Using the above relation in equation \eqref{PBH4} we get the vacuum field equations as,
\begin{equation}\label{PBH7}
	R_{\alpha\beta}=\frac{R_0}{4}g_{\alpha\beta}.
\end{equation}

Equation \eqref{PBH7} represents a constant curvature space { which predicts a Schwarzchild-de Sitter black hole when $R_0/4$ is considered to be the cosmological constant. $R=0$ gives standard Schwarzschild black hole.} Therefore, $f(R)$ gravity generally permits a black hole solution with a cosmological background and particularly permits a Schwarzschild black hole. 

%If one identifies $R_0/4$ as the cosmological constant then Schwarzschild-de Sitter black hole is a natural outcome of the theory. If $R_0=0$, one deduces the standard Schwarzschild black hole solution. Therefore, $f(R)$ gravity generally permits a black hole solution with a cosmological background and particularly permits a Schwarzschild black hole.

\subsection{Scalarons and primordial black holes}

Black holes in GR, which are stationary and end products of gravitational collapse are described by the Kerr-Newman metric. These are uniquely described by three parameters --- mass, spin and charge and { no additional degrees of freedom are required to describe the spacetime structure of these objects}. These are regarded as the simplest objects in the universe. It is known as the `no-hair' theorem. \cite{1972CMaPh..25..152H} demonstrated that black holes in GR and Brans-Dicke theory of gravity are indistinguishable from each other. This idea was further generalised by \cite{PhysRevLett.108.081103} to more general class of scalar-tensor theories including $f(R)$ theories by assuming stationary condition and neglecting matter distribution { and} it provided with a strong theoretical basis for the `no-hair' theorem.

However, the idea of hairy black holes has gained momentum in the new era of dark matter physics and observational GR. \cite{PhysRevLett.83.2699} studied evolution of primordial black holes in scalar-tensor cosmology and showed that a minimally coupled scalar field with non-trivial time variation endows the black hole with scalar hair. Generalised scalar-tensor theories permit black hole hair \citep{PhysRevLett.108.081103, 2015CQGra..32u4002S}. Asymptotically flat black hole solution with scalar hair has been reviewed in \cite{2015IJMPD..2442014H}. Growth of non-trivial scalar field profile around Schwarzschild black holes has been demonstrated in \cite{PhysRevD.100.063014} with the consideration of the black hole embedded in a background of time varying and homogeneous scalar field. Testability of scalar field conformally coupled to curvature and scalar fields with minimal coupling with a potential through shadow cast by M87* supermassive black hole has been reported by \cite{2020JCAP...09..026K}. \cite{2019JCAP...06..038H} showed that scalar dark matter around a black hole develops scalar hair. Scalar field profile near a black hole has been found to be dependent on the scalar particle mass. While the field profile for low mass scalars ($<<10 ^{ -20}$ eV) is $1/r$ at large scale, it is $1/r^{3/4}$ for high mass scalars ($>>10^{ - 20}$ eV). The Compton wavelength of a 10$^{-20}$ eV scalar particle corresponds to the size of the Schwarzschild radius of the M87* black hole. Therefore, dark matter profile near the horizon gives information about the dark matter particle mass { which} is of potential influence on the well recognised riddle of black hole information loss paradox.

Here we consider the $f(R)$ gravity degree of freedom --- the scalarons, which are of gravitational origin near the black hole horizon. The Lagrangian $f(R)$ in the Einstein-Hilbert action (equation \eqref{PBH1}) appears naturally through curvature corrections to the quantum vacuum fluctuations \citep[see \citep{Kalita_2018, 2020ApJ...893...31K} for details]{Sakharov:1967pk, ruzmaikina1970quadratic}. The scalaron field $\psi=f^\prime(R)$ is determined by the ultraviolet (UV) and infrared (IR) scales of vacuum fluctuations { where} the UV wave number is chosen as $k_\text{UV}=2\pi/R_\text{s}$ with $R_\text{s}$ being the Schwarzschild radius { and} the IR wave number is chosen as $k_\text{IR}=2\pi/\lambda_\text{IR}$ with the IR scale $\lambda_\text{IR}$ corresponding to thermal energy density in vacuum, $u_{\text{th}}=a_BT_\text{H}^4={hc}/{\lambda_\text{IR}^4}$. Here $T_\text{H}$ is the Hawking temperature \citep{HAWKING1974} of the black hole of mass ${M}$,
\begin{equation}\label{PBH8}
	T_\text{H}=\frac{\hbar c^3}{8\pi k_BGM}.
\end{equation}

Considering a linear perturbation of the scalaron field around a background curvature $R_0$ near a black hole, \cite{2020ApJ...893...31K} constructed a dimensionless measure of the scalaron field amplitude $(\psi_0)$ and deduced the scalaron mass $(m_\psi)$ in terms of the vacuum fluctuation scales. These are expressed as follows,

\begin{equation}\label{PBH9}
	\psi_0=\frac{2\pi}{k_BT_\text{H}}\left({\frac{1}{k_\text{UV}k_\text{IR}}}\right)^{1/2}\left(\frac{\hbar c}{2}\right)\left(k_\text{UV}^2-k_\text{IR}^2\right),
\end{equation}

\begin{equation}\label{PBH10}
	m_\psi=\left[{\frac{k_\text{UV}^2-k_\text{IR}^2}{12ln\left(\frac{k_\text{UV}}{k_\text{IR}}\right)}}\right]^{1/2} \Biggl(\frac{h}{c} \Biggr). 
\end{equation}

The scalaron mass is inversely proportional to the mass of the black hole (see equations \eqref{PBH11}), with the proportionality constant completely determined by the fundamental constants of gravity, quantum theory and thermodynamics. 
\begin{subequations}
	\label{PBH11}
	\begin{equation}\label{PBH11a}
		m_\psi=f(G, c, h, a_B, k_B; {M}_{\odot}) \biggl(\frac{{M}}{{M}_{\odot}} \biggr)^{-1} ,
	\end{equation}
	\text{where,}
	\begin{equation}\label{PBH11b}
	   f=\Biggl( \dfrac{h}{c}\Biggr) \left[\frac{\pi^2c^4}{G^2}-\frac{\left(a_Bh^3c^3\right)^{1/2}c^4}{\left(8\pi Gk_B\right)^2}\right]^{1/2}\left[\frac{1}{12ln\left(\frac{8\pi^2k_B}{\left(a_Bh^3c^3\right)^{1/4}}\right)}\right]^{1/2} \Biggl(\frac{1}{{M}_{\odot}}\Biggr).
	\end{equation}
\end{subequations}

{ In the unit of energy (eV), the scalaron mass is expressed as,
\begin{equation}
	m_\psi=4.30\times 10^{-10} \text{ eV} \left( \frac{M}{M_\odot}\right)^{-1} 
\end{equation}
}

This relationship between scalaron mass and black hole mass is a consequence identical to the scenario of formation of black holes through gravitational collapse of self interacting, minimally coupled scalar fields studied by \cite{2018EPJC...78..676C}. These scenarios are particularly appealing when the scalar field is an axion-like field. Axions are alternatives to WIMPs as dark matter candidates. Axionic black holes are potential targets for future gravitational wave observatories \citep{PhysRevD.91.084011}.

\begin{figure}
\centering
\includegraphics[scale=0.85]{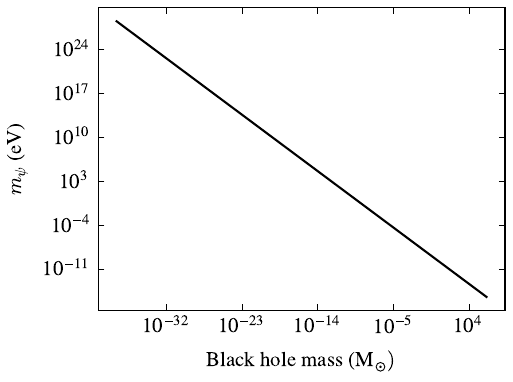}
\includegraphics[scale=0.85]{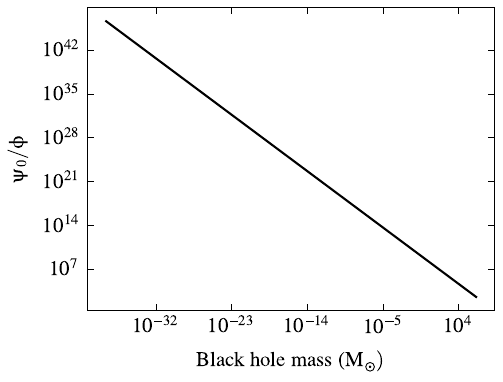}
\caption{\emph{Left :} Scalaron mass vs PBH mass,
		\emph{Right :} \emph{LPTK-Psi-Phi} plot for various PBH masses}
\label{fig1}
\end{figure}

WIMP class $(100-1000 \text{ GeV})$ scalarons require $10^{ - 21} -10 ^{- 22}\text{ } {M}_{\odot}$ PBHs { whereas} axion-like particles $({10}^{-5}\text{ eV}\le m_\psi\le 1$ eV) require PBHs with masses ranging from $10 ^{- 5}\text{ } M_{\odot}$ (Uranus like) to $10^{ - 10}\text{ } M_{\odot}$ (Enceladus like). If scalarons are like ultralight fuzzy dark matter $(m_\psi\approx{10}^{-22}$ eV) PBHs of $10^{12}\text{ }M_{\odot}$ are required. This relationship is depicted in figure~\ref{fig1} (\emph{Left}).

To constrain scalaron mass ($f(R)$ modified gravity theory) and hence the PBH mass one has to examine its correction to known physical processes occurring in the framework of GR. In our case the processes constitute the BBN. Any modification to GR has to be screened in environments where this theory has been successfully tested \citep{2015PhR...568....1J, 2019ARA&A..57..335F}. BBN with standard general relativistic field equations of cosmic expansion has been a successful paradigm{, and} therefore, putting constraints on $f(R)$ gravity scalarons through the BBN calls for testing unscreening of the scalaron degree of freedom in that epoch. Several screening mechanisms exist for suppressing modified gravity degrees of freedom so that available local tests of GR are not spoiled \citep{PhysRevD.76.064004, Starobinsky_2007}. One such mechanism is the chameleon mechanism \citep{ PhysRevD.69.044026, PhysRevLett.93.171104} { in which} extra degrees of freedom are screened in high density environments and get unscreened in low density environments. If $\psi_0$ is the dimensionless scalaron field amplitude and $\phi=GM/c^2l$ is the dimensionless gravitational potential on a scale $l$ then ${\psi_0}/{\phi}>1$ gives unscreened scalarons { whereas ${\psi_0}/{\phi}<1$ indicates that scalarons are screened}. We choose the scale $l$ as the particle horizon, $2ct$ at the  onset of BBN, $t =1$ s { which} is around the solar radius $(\approx 0.004$ au). For calculating scalaron field amplitude through equation \eqref{PBH9} we consider the PBH mass range $10^{ - 38} - 10^6\text{ }M_{\odot}$. This sets the UV and IR scales. The `Psi-Phi' plot (as we call it in this paper) is shown in figure~\ref{fig1} (\emph{Right}). We wish to name it as \emph{LPTK-Psi-Phi} plot for the authors \emph{Lalremruati} (see \citep{2022ApJ...941..183L} for the use of `Psi-Phi' plot in the investigation of effect of dark matter on screening of scalarons near the Galactic Centre black hole), \emph{Paul} (see \citep*{2023IJMPD..3250021P} who investigated testability of unscreened scalarons near the orbit of the S-2 star orbiting the Galactic Centre black hole), \emph{Talukdar} (one of us in this paper who generated this plot for PBHs) and \emph{Kalita} (see \citep{2020ApJ...893...31K, 2021ApJ...909..189K} for first reports of testability of scalarons near the Galactic Centre black hole through existing and upcoming astrometric facilities employed for measuring pericentre shift of compact stellar orbits near the black hole). It is seen that scalarons corresponding to the above considered PBH mass range are all unscreened in the BBN era { which} indicates that there is possibility to constrain them and thereby the PBH masses through their influence on the BBN process.

\section{BBN Test}\label{sec3}

BBN is a strong probe for constraining new physics. Using BBN as a probe, several attempts have been made in the past to constrain modified gravity models \citep{Casas:1990fz, Anagnostopoulos2022gej, Lambiase:2012fv,Capozziello:2017bxm, Sultan:2022aoa, Asimakis:2021yct, Asimakis:2022kfk,Bhattacharjee:2021hwm}, number of particle species \citep{Ichikawa:2007fa,PhysRevD.97.023502,Yeh:2022heq} and dark energy models \citep{PhysRevD.67.063501}. In this section we produce constraint on the scalaron mass through BBN and hence generate allowed PBH mass range { using} the observed shift of freezeout temperature and elemental abundance of $^4$He nuclei.

The process of synthesis of light elements starting from hydrogen ($^1$H) up to $^7$Li in the early universe is well known. It incorporates the rate of weak interaction among the elementary particle species ($n+\nu_e \rightleftharpoons p + e^{-}$,  $n+e^{+} \rightleftharpoons p + \bar{\nu_e}$,  $n \rightleftharpoons p + e^{-}+\bar{\nu_e}$) and the cosmic expansion rate predicted by standard Friedmann equation of GR. These weak processes are responsible for neutron to proton conversion and vice-versa. The weak interaction rate ($\lambda$) and cosmic expansion rate ($H$) vary with temperature as $T^5 \text{ and } T^2$ respectively{, and therefore}, as the universe expands, the weak interaction rate falls faster than the expansion rate and an epoch comes when they become equal to each other. This epoch is known as the  `freezeout epoch'.

Modification to the theory of gravity elevates the expansion rate of the universe \citep{PhysRevD.70.043528,Capozziello:2002rd, Capozziello:2003gx,PhysRevD.68.123512, 2007IJGMM..04..115N} { which} causes the freezeout to occur at a temperature higher than that predicted by GR. The { shift of freezeout epoch to earlier time} gives a different neutron to proton ($n_n/n_p$) ratio { which} in turn changes the primordial $^4$He abundance. This principle has been employed to constrain the mass of $f(R)$ gravity scalarons and and hence the PBH mass range.

Modification of gravity can take into account the phenomena attributed to the presence of non-baryonic dark matter \citep{Nojiri:2008nt}. It has been shown that scalaron may act as dark matter candidate in Starobinsky $f(R)$ gravity theory \citep{Cembranos:2008gj, Katsuragawa:2017wge, Katsuragawa:2016yir}. {In the present study we see }that $f(R)$ scalarons are unscreened across the particle horizon at the onset of BBN ($t \approx1$ s) (see section \ref{sec2}). 

{
In $f(R)$ gravity theory the Friedmann equation for cosmic expansion in the radiation era is expressed as \citep{Sotiriou:2008rp},
\begin{equation}
	H^2=\left(\frac{\dot{a}}{a}\right)^2=\frac{8\pi G}{3}\left(\frac{\rho_r}{f^\prime(R)}+\rho_{f(R)}\right)
\end{equation}

where  $\rho_r$ is the background radiation density and $\rho_{f(R)}$ is the effective mass density of geometry defined as,

\begin{equation}
	\rho_{f(R)}=\frac{Rf^\prime\left(R\right)-f(R)}{f^\prime(R)}-\frac{3H\dot{R}f''(R)}{f^\prime(R)}
\end{equation}

The main motivation for defining an effective density of geometry is to use $f(R)$ gravity in explaining accelerated expansion of the universe without resorting to exotic negative pressure fields or fluids. The scalaron field is defined as $\psi=f^\prime\left(R\right)$. With this the Friedmann equation can be written as,

\begin{equation}
	H^2=\frac{8\pi G}{3}\left(\rho_r\left(\frac{1-\psi}{\psi}\right)+\rho_r+\rho_{f(R)}\right)
\end{equation}
	
If one assumes only small departure from general relativity the scalaron field amplitude is very close to unity, $\psi\approx1$. Under this condition Friedmann equation takes the form,

\begin{equation}
	H^2=\frac{8\pi G}{3}\left(\rho_r+\rho_{f(R)}\right)
\end{equation}

To make the analysis model independent and extract the information of the scalarons we identify scalarons as the additional source term in the Friedmann equation and identify $\rho_{f(R)}$ as the scalaron mass density, $\rho_{m_\psi}$. This is justified from another point of view – that modification to gravity and modification to matter side of Einstein’s field equations are one and the same physics \citep{maartens_durrer_2010}.} {Therefore, we treat scalarons as additional degrees of freedom which contributes to a density component affecting the cosmic expansion rate.} To investigate the effects of $f(R)$ scalarons in the BBN era, we consider the Friedmann equation in the modified gravity scenario as,  
\begin{equation}\label{BBN1}
H=\sqrt{\dfrac{8\pi G}{3} (\rho_{m_{\psi}}+\rho_r)},
\end{equation}
where, $\rho_{m_{\psi}}=n_{\psi}m_{\psi}$ is mass density of the scalarons with $m_{\psi}$ and $n_{\psi}$ being the mass and number density of the scalarons respectively. $\rho_r$ is the mass density of the relativistic species.

%The modified Friedmann equation is then written as,

%\begin{equation}
	%H^2=\frac{8\pi G}{3}\left(\rho_r+\rho_{m_\psi}\right)
%\end{equation}

From the above equation, we obtain the shift of the expansion rate as,
\begin{equation}\label{BBN2}
	\delta H= H_\text{GR} \left( \sqrt{1+\dfrac{\rho_{m_{\psi}}}{\rho_r}}-1\right).
\end{equation}

The weak interaction rate among the particle species is given by \citep{Weinberg:1972kfs},

\begin{equation}\label{BBN3}
	\lambda=\dfrac{7\pi^{5}}{15}\left( \frac{g_\nu^2+3 g_A^2}{2\pi^3 \hbar^7}\right)  (k_B T)^5 c^{-6}, 
\end{equation}
where, 
$g_\nu$ and $g_A$ are weak interaction coupling constant and axial vector coupling constant given by, $g_\nu=1.418 \times 10^{-62} \text{ J m}^3$ and $g_A=1.18g_\nu=1.673 \times 10^{-62}\text{ J m}^3$ respectively. Here $c$ represents the velocity of light in vacuum. We rewrite equation \eqref{BBN3} as,
\begin{equation}\label{BBN4}
	\lambda= F_q T^5,
\end{equation}
where, $F_q \approx (1.1347 \times 10^{-50})$ K$^{-5}$ s$^{-1}$. It gives shift of interaction rate,
\begin{equation}\label{BBN5}
	\delta \lambda=5F_q T^4 \delta T.
\end{equation}

In the freezeout epoch, we have $\lambda=H$ { which} corresponds to the standard freezeout temperature in GR, $T_f \approx 9 \times 10^{9} \text{ K}$. The shift of the freezeout temperature is obtained from
\begin{equation}\label{BBN6}
		\delta \lambda=\delta H .
\end{equation}

We assume that the mass density of scalarons { to be} smaller than the mass density of standard relativistic species ($\rho_{m_{\psi}}<<\rho_r$). Putting the expressions for $\delta\lambda$ (equation \eqref{BBN5}) and $\delta H$ (equation \eqref{BBN2}), we have with the above assumption the following expression for the shift of freezeout temperature,
\begin{equation}\label{BBN7}
	{\dfrac{{\delta T_f}}{{T_f}}=\dfrac{{\rho_{{m_{\psi}}}}}{{\rho_r}}	\dfrac{{H_\text{GR}(T_f)}}{{10F_qT_f^5}}},
\end{equation}
where, $H_\text{GR}(T_f)$ is the general relativistic expansion rate in the freezeout epoch.

Number density of bosonic species in thermal equilibrium is given as,
\begin{equation}\label{BBN8}
	n=\dfrac{g}{2\pi^2 \hbar^3}\int_0^\infty \dfrac{p^2 dp}{\text{e}^{E/k_B T}-1}.
\end{equation}

For scalarons, spin degree of freedom is, $g=g_{\psi}=1$. Considering non-relativistic scenario ($E={p^2}/{2m}$) , we obtain the number density as (in m$^{-3}$),
\begin{equation}\label{BBN9}
	n = n_{\psi} \approx 7.26 \times 10^{66} \text{ }m_{\psi}^{{3}/{2}}T^{{3}/{2}}. 
\end{equation}

Therefore, the mass density of the scalarons becomes (in kg m$^{-3}$),
\begin{equation}\label{BBN10}
	\rho_{m_\psi}=m_{\psi}n_{\psi} \approx 7.26 \times 10^{66} \text{ }m_{\psi}^{{5}/{2}}T^{{3}/{2}}. 
\end{equation}

To calculate $H_\text{GR}(T_f)$, we have used $\rho_r={N a_B T_f^4}/{c^2}$ where, $N=(g_b/2)+({7}/{16})g_f\left({4}/{11}\right)^{4/3}=1.68$ in case of standard particle species in the Big Bang model, $g_b=2$ (for photons) and $g_f=6$ (for 3 neutrino species). Radiation mass density at freezeout is calculated as,
\begin{equation}\label{BBN11}
	\rho_r(t_f) \approx 9.27 \times 10^7 \text{ kg m$^{-3}$}.
\end{equation}

The general relativistic expansion rate is obtained as,
\begin{equation}\label{BBN12}
	H_\text{GR}(T_f) \approx 0.22 \text{      s$^{-1}$}.
\end{equation}

Substituting equation \eqref{BBN10}, \eqref{BBN11} and \eqref{BBN12} in \eqref{BBN7}, we obtain the shift in freezeout as, 
\begin{equation}\label{BBN13}
	\left( \dfrac{\delta T_f}{T_f}\right)^{\text{NRel}}  \approx (2.27 \times 10^{72}) m_{\psi}^{5/2} { \text{ kg}^{-5/2}},
\end{equation} 

{
For relativistic scalarons, $E \approx pc$. Following the method for non-relativistic scalaron, the shift in freezeout temperature has been obtained as, 

\begin{equation}
	\left( \frac{\delta T_f}{T_f}\right)^\text{Rel} \approx (2 \times 10^{27}) m_\psi \text{  kg}^{-1}
\end{equation}

Here, $m_{\psi}$ is the scalaron mass in kg.
}

The abundance of primordial $^4$He is written in terms of neutron fraction ($X_n={n_n}/({n_n+n_p})$) as \citep{PhysRevLett.79.1588},
\begin{equation}\label{BBN14}
	Y_p=2X_n(t_f)=2\left[\dfrac{1}{1+\exp(Q/k_B T_f)} \right] \left[\exp\left(-\dfrac{t_{nuc}-t_f}{\tau_n}\right)\right]=C \dfrac{2x(t_f)}{1+x(t_f)},
\end{equation}
where,  $t_f$ is the freezeout time in GR ($\approx1$ s),
$t_{nuc}$ is the deuterium synthesis time ($\approx$ $200$ s),
$\tau_n$ is the neutron $\beta$-decay life time ($\approx$ 877.75 s \citep{doi:10.1126/science.aan8895}), $Q$ is the neutron-proton mass difference ($\approx$ 1.293 MeV = 2.06 $\times 10^{-13}$ J),
$x(t_f) = \exp\left( -{Q}/{k_B T_f}\right) $ and $C=\exp\left[-({t_{nuc}-t_f})/{\tau_n}\right]$.

\subsection{Shift of freezeout in \boldmath$f(R)$ gravity}

Differentiating equation \eqref{BBN14} and using time-temperature relation $t_f \varpropto 1/T_f^2$, the shift in $^4$He abundance is obtained as \citep{PhysRevLett.79.1588},

\begin{equation}\label{BBN15}
	{\dfrac{{\delta Y_p}}{{Y_p}}=\left[ \left( 1-\dfrac{{Y_p}}{{2C}}\right){ln}\left( \dfrac{{2C}}{{Y_p}}-1\right)-\left( \dfrac{{2t_f}}{{\tau_n}}\right)   \right] \dfrac{{\delta T_f}}{{T_f}}}.
\end{equation}

In general relativistic BBN history $Y_p \approx 0.26$. Putting the value of other associated numbers in equation \eqref{BBN15}, we obtain,

\begin{equation}\label{BBN16}
	\dfrac{\delta Y_p}{Y_p} \approx 1.39 \left(\dfrac{\delta T_f}{T_f}\right). 
\end{equation}

Upper bound on the shift of freezeout temperature and $^4$He abundance is employed to constrain the mass of scalarons (see equations \eqref{BBN13} and \eqref{BBN16}). We take the upper bound on the observed value of shift of freezeout temperature and $^4$He abundance given by \cite{Lambiase:2012fv},
\begin{equation}\label{BBN17}
	\begin{aligned}
		\left|\dfrac{\delta T_f}{T_f}\right| \le& 4.7 \times 10^{-4},\\
		\left|\dfrac{\delta Y_p}{Y_p}\right| \le& 4.04 \times 10^{-4}.
	\end{aligned}
\end{equation}

\begin{figure}
\centering
\includegraphics[scale=0.85]{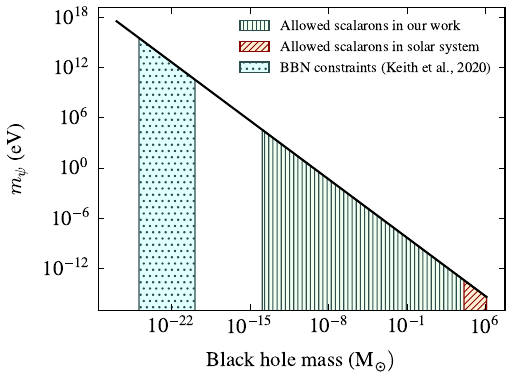}
~
\includegraphics[scale=0.85]{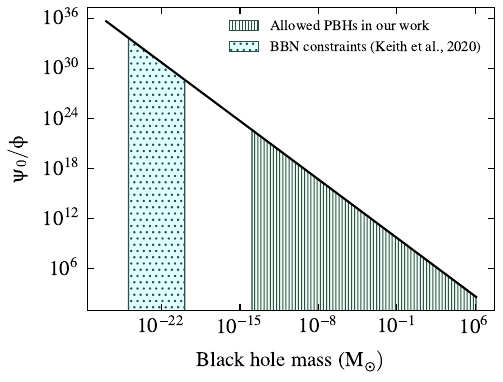}
\caption{\emph{Left :} Scalaron mass vs PBH mass obtained from BBN constraint,
		\emph{Right :} \emph{LPTK-Psi-Phi} plot for constrained PBH masses}
\label{fig2}
\end{figure}

    This produces allowed range of {non-relativistic} scalaron mass as $10^{-16}-10^4\text{ eV}$. The corresponding PBH mass range (see equations \eqref{PBH11}) is found to be $10^{6}-10^{-14}\text{ }M_{\odot}$. { Similarly, the allowed range of mass of the relativistic scalarons is also found to be $10^{-16}-10^4\text{ eV}$ which produces the same PBH mass range derived from non-relativistic scalarons. It is to be noted that the upper bound on scalaron mass ($\approx 10^4 \text{ eV} $) derived in this way is far below $10^{28} \text{ eV}$ which corresponds to scalarons associated with Planck mass PBHs ($M=10^{-8}$ kg)}. The variation of scalaron mass with constrained PBH masses is shown in figure~\ref{fig2} (\emph{Left}). The PBH mass range $10^{-25}-10^{-20}\text{ }M_{\odot}$ was previously found to be prohibited from the consideration of synthesis of D and $^4$He \citep{PhysRevD.102.103512}. This mass range is also superimposed in figure~\ref{fig2} (\emph{Left}). PBHs with masses in the range $10^{-24}-10^{-19}\text{ }M_{\odot}$ undergo significant evaporation during BBN, thereby affecting the primordial abundances of elements. Therefore such mass range is highly constrained \citep{Dienes:2022zgd}. But, PBHs with $M< 10^{-24}M_{\odot}$ completely evaporates before BBN { and} they remain unconstrained through the Hawking evaporation process. However, very small PBHs have implication for early universe cosmology such as the production of dark matter and dark radiation \citep{2019JHEP...08..001H, 2020EPJP..135..552M, 2020JCAP...08..045B} and baryogenesis \citep{PhysRevD.43.984, 2017PTEP.2017c3B02H, PhysRevD.103.043504}. In this situation, scalarons serve as a probe to constrain those small PBHs. The allowed range of PBH masses found in our work is well above the mass window which spoils the light element synthesis. The \emph{LPTK-Psi-Phi} plot for the constrained range of PBH masses is shown in figure~\ref{fig2} (\emph{Right}) { from which it is clearly evident} that the scalarons are unscreened even for constrained PBH masses. 

In local tests of $f(R)$ gravity, scalaron mass enters into the PPN parameter $\gamma$. In the weak and static field limit of $f(R)$ gravity theory this parameter is given by \citep{Kalita_2018},
\begin{equation}
	\gamma=\dfrac{3-\text{e}^{-m_{\psi}l}}{3+\text{e}^{-m_{\psi}l}},
\end{equation} 
where $l$ is the scale of a system where the test is performed. 

The observed bound on $\gamma-1$ obtained from solar system experiments such as the Cassini mission \citep{2003Natur.425..374B} is realised as $\gamma-1 \approx -10^{-5}$. This gives $m_{\psi}l \approx 11.11$. For $l \approx 0.004$ au (solar radius) and $l \approx 800$ au (expected distance to the hypothetical `Planet 9'), we obtain the solar system constraint on scalaron mass as (it is to be noted that $13.08 \text{ au}^{-1}=1.07 \times 10^{-16}\text{ eV}$),
\begin{equation}
	m_{\psi} \approx 10^{-19}-10^{-14} \text{  eV}.
\end{equation}

The window $10^{-16}-10^{-14}$ eV { of scalaron mass is a segment of the mass range obtained from solar system constraint. The solar system constraint} is superimposed on our allowed range obtained from BBN in figure~\ref{fig2} (\emph{Left}).

\subsection{Neutron fraction, deuterium fraction and (D+\boldmath$^3$He) abundance with scalarons}

%\subsubsection{Modified time temperature relation}

Neutron fraction and temperature are essential parameters for calculation of deuterium mass fraction and hence, the baryonic matter content of the universe. Variation of both of them depends on the expansion rate governed by { the} theory of gravity. The time-temperature relation in GR based Friedmann model is given as,
\begin{equation}\label{BBN18}
	T_\text{GR}=\left( \dfrac{3c^2}{32\pi G a_B}\right)^{1/4}N^{-1/4}t^{-1/2} \approx \dfrac{1.5 \times 10^{10}\text{ K}}{\sqrt{t\text{ (sec)}}} ,
\end{equation}
where, $T_\text{GR}$ is the temperature in the standard Friedmann model corresponding to time $t$. The neutron fraction above the freezeout temperature in GR is given by the equilibrium expression \citep{Weinberg2008},
\begin{equation}\label{BBN19}
	X_n(t)=\left[\dfrac{1}{1+\exp(Q/k_B T_\text{GR})} \right].
\end{equation}

For temperature below the general relativistic freezeout temperature ($T < T_f^\text{GR}$), we have, neutron fraction given by the $\beta$-decay expression as,
\begin{equation}\label{BBN20}
	X_n(t)=\left[\dfrac{1}{1+\exp(Q/k_B T_f^\text{GR})} \right]\left[\exp\left(-\dfrac{t-t_f^\text{GR}}{\tau_n}\right)\right],
\end{equation}
where, $t_f^\text{GR}$ is the time corresponding to $T_f^\text{GR}$. Now we consider relativistic and non-relativistic scalaron. For relativistic scalarons, $\rho_{{m_{\psi}}}={g_{\psi}a_B T^4}/{2c^2} $ (where $g_\psi=1$ is the spin degeneracy factor for scalarons). This gives us the following relation,
\begin{equation}\label{BBN21}
	\dfrac{\rho_{{m_{\psi}}}}{\rho_r}=\dfrac{1}{2}.
\end{equation}

On the other hand, for non-relativistic scalarons, we use equation \eqref{BBN10} and obtain the following relation,
\begin{equation}\label{BBN22}
	\dfrac{\rho_{{m_{\psi}}}}{\rho_r}=5.15 \times 10^{98} \text{ } m_\psi^{5/2} \text{ } \left( {T_{f(R)}^\text{NRel}}\right) ^{-5/2},
\end{equation}
where, ${T_{f(R)}^\text{NRel}}$ is the temperature in the non-relativistic $f(R)$ gravity scenario.

We derive the modified time-temperature relations for relativistic and non-relativistic  scalarons as follows, 

The modified Friedmann equation in presence of scalaron written as,
\begin{equation}\label{BBNFRD1}
	\dfrac{1}{a^2}\left( \dfrac{da}{dt}\right)^2=\dfrac{8\pi G \rho_r}{3}\left( 1+ \dfrac{\rho_{{m_{\psi}}}}{\rho_r}\right) . 
\end{equation}

For relativistic scalarons, we substitute equation \eqref{BBN21} and time-scale factor relation in radiation dominated era $T \varpropto (1/a)$ in the above equation and obtain time-temperature relation,
\begin{equation}\label{BBN23}
	T_{f(R)}^\text{Rel}=\dfrac{7}{8}\text{ }T_\text{GR}.
\end{equation}

For non-relativistic scalarons, we rewrite ${\rho_{{m_{\psi}}}}/{\rho_r}$ (equation \eqref{BBN22}) in terms of cosmic scale factor $a$ as,
\begin{equation}\label{BBNFRD2}
	\dfrac{\rho_{{m_{\psi}}}}{\rho_r}=4.18 \times 10^{97} \text{ }m_{\psi}^{5/2} a^{5/2}.
\end{equation}

Substituting the above expression in modified Friedmann equation (see equation \eqref{BBNFRD1}) and assuming the condition ${\rho_{{m_{\psi}}}}<<{\rho_r}$, we obtain the following relation,
\begin{equation}\label{BBNFRD3}
	\bigints \dfrac{ada}{[1+( 4.18 \times 10^{97} \text{ }m_{\psi}^{5/2} a^{5/2}) /{2}]}=\sqrt{\dfrac{8 \pi G \rho_{r(0)}}{3}} \bigints dt.
\end{equation}

Considering the upper and lower bound on the scalaron mass obtained through BBN ($m_\psi^{\text{Lower}}=6.67 \times 10^{-52}$ kg ($\approx 10^{-16}$ eV) and $m_\psi^{\text{Upper}}=6.67 \times 10^{-32}$ kg ($\approx 10^4$ eV) respectively) and using them in equation \eqref{BBNFRD3}, we obtain,
\begin{subequations}\label{BBNFRD4}
\begin{equation}\label{BBNFRD4a}
	T_{f(R)}^\text{NRel}=T_\text{GR} \left( \dfrac{1}{\, _2F_1\left(0.8,1;1.8;-2.4 \times 10^{-31}a^{5/2}\right)} \right)^{-1/2}
	\text{   , for $m_\psi^{\text{Lower}}$}
\end{equation}
\text{and}
\begin{equation}\label{BBNFRD4b}
	T_{f(R)}^\text{NRel}=T_\text{GR} \left( \dfrac{1}{\, _2F_1\left(0.8,1;1.8;-2.4 \times 10^{19}a^{5/2}\right)} \right)^{-1/2}
	\text{ , for $m_\psi^{\text{Upper}}$}
\end{equation}
\end{subequations}

The first order hypergeometric functions ($_2F_1$) appearing in equations \eqref{BBNFRD4} approach unity for small scale factor regime , $a=a_\text{BBN} \approx 10^{-10}$. Hence, we retain the GR time-temperature relation for non-relativistic scalarons,
\begin{equation}\label{BBN25}
	T_{f(R)}^\text{NRel}=T_\text{GR}.
\end{equation}

The elevated freezeout in $f(R)$ gravity is calculated as follows. We have the freezeout temperature in GR as $T_{f}^\text{GR}=9 \times 10^9 \text{ K}$. Modification in the Friedmann equation increases the expansion rate and hence the freezeout temperature is elevated. For the freezeout epoch we equate the expansion rate (equation \eqref{BBN1}) and the rate of interaction among the particle species (equation \eqref{BBN4}) and deduce the elevated freezeout temperature as,

\begin{equation}\label{BBN27}
	T_f^{f(R)}=T_f^\text{GR}\left(1+\dfrac{1}{6} \dfrac{\rho_{{m_{\psi}}}}{ \rho_r} \right). 
\end{equation}

This is a generalised expression for modified freezeout temperature for both relativistic and non-relativistic scalarons. Therefore, their values are,
\begin{equation}\label{BBN28}
	\begin{aligned}
		&T_f^{f(R)_\text{Rel}}=9.75 \times 10^9 \text{ K}, \\
		&	T_f^{f(R)_\text{NRel}}=9 \times 10^9 \text{ K}.		
	\end{aligned}
\end{equation}

\begin{figure}
\centering
\includegraphics[scale=0.85]{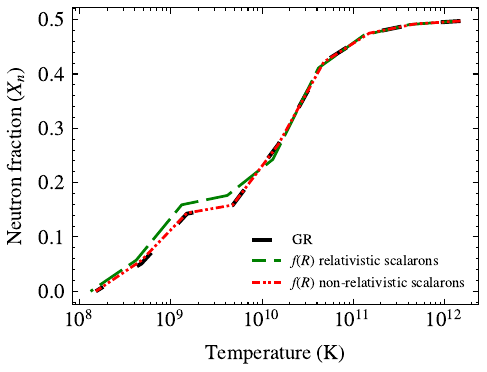}
~
\includegraphics[scale=0.85]{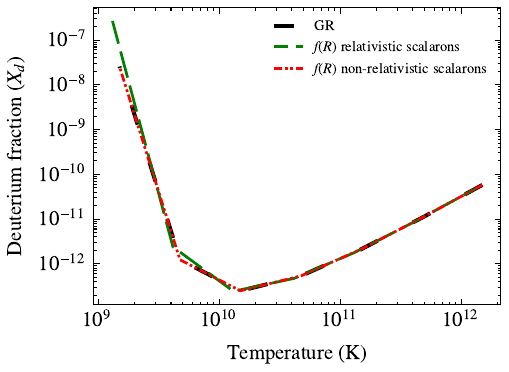}
\caption{\emph{Left :} Variation of neutron fraction with temperature in GR and $f(R)$ gravity,
		\emph{Right :} Variation of deuterium fraction with temperature in GR and $f(R)$ gravity (for both relativistic and non-relativistic scalaron)}
\label{fig3}
\end{figure}

It is seen that the departure of freezeout temperature from that predicted by GR based expansion rate is small. For a range of temperature, the variation of neutron fraction ($X_n$) for GR and $f(R)$ are shown in figure~\ref{fig3} (\emph{Left}). The modified time-temperature relations help us to calculate deuterium fraction at various temperatures relevant to the BBN epoch.

Synthesis of deuterium is an important event in BBN history. It marks the onset of production of $^4$He and $^7$Li in the primordial universe. Following \cite{Weinberg2008}, deuterium fraction is written as,
\begin{equation}\label{BBN29}
	X_d=\dfrac{n_d}{n_b}=3\sqrt{2}X_pX_n\epsilon \exp\left( {\dfrac{B_d}{k_B T}}\right). 
\end{equation}

Here, $n_d$ is the number density of deuterium, $n_b$ is the number density of baryons ($n_p+n_n$), $T$ is the temperature at which $X_d$ is being calculated, $X_n$ and $X_p=1-X_n$ are neutron and proton mass fractions evaluated at temperature $T$, $B_d$ is the deuterium binding energy ($\approx 2.2 $ MeV) and $\epsilon$ is a dimensionless quantity given by,
\begin{equation}\label{BBN30}
	\epsilon \approx 1.46 \times 10^{-12} \left( \dfrac{T}{10^{10} \text{K}}\right)^{3/2} \Omega_B h^2. 
\end{equation}

We take the present baryon density parameter as $\Omega_B \approx 0.02 h^{-2}$. The deuterium fraction is not appreciable until the temperature comes down to $T_d \approx 7 \times 10^8 \text{ K}$. The exponential factor in equation \eqref{BBN29} is elevated and it suppresses the smallness of $\epsilon$. Taking into account the time-temperature relations for relativistic and non-relativistic scalarons (see equations \eqref{BBN23} and \eqref{BBN25}) the variation of deuterium fraction ($X_d$) with temperature is shown in figure~\ref{fig3} (\emph{Right}).

Deuterium (D) and $^3$He are burned into $^4$He in the BBN process through two body reactions. A large value of baryon-to-photon ratio ($\eta={n_b}/{n_{\gamma}}$) reduces D and $^3$He abundance as the two body reaction rate is proportional to $\eta$. A competition between two body reactions converting D and $^3$He into $^4$He and the expansion rate governs the final abundance of D and $^3$He \citep{1985ARA&A..23..319B}, { and therefore,} a modification to theory of gravity has well defined influence on the combined abundance  $({\text{D}+{^3}\text{He}})/{\text{H}}$. 

The scalarons elevate the cosmic expansion rate due to which the timescales are shortened. Following \cite{1985ARA&A..23..319B}, the parametrisation of shortening of timescale is given by,
\begin{equation}
	t'=\zeta^{-1}t, \text{  $\zeta \ge 1$},
\end{equation}
where, $t'$ and $t$ represent time in $f(R)$ gravity and GR respectively. $\zeta$ is known as `speedup factor' { which} encapsulates the modification to expansion rate. Replacing the time by Hubble rate, we get,
\begin{equation}\label{DH2}
	\zeta=\dfrac{t}{t'}=\dfrac{H}{H_\text{GR}}=\sqrt{1+\dfrac{\rho_{{m_{\psi}}}}{\rho_r}},
\end{equation}
where, $H$ and $H_\text{GR}$ are the Hubble rates in $f(R)$ scalaron gravity and GR respectively.

The combined D and ${^3}$He abundance is given by \citep{1985ARA&A..23..319B},
\begin{equation}\label{DH3}
	\dfrac{\text{D}+{^3}\text{He}}{\text{H}}=5 \times 10^{-4}\left( \dfrac{\zeta}{\eta_{10}}\right)^{1.4} ,
\end{equation}
where, $\eta_{10}=\eta/10^{-10}$. Here we use the latest Planck data for the present value of $\eta_{10}$ \citep{Yu:2021djs},
\begin{equation}\label{DH4}
	\eta_{10}=6.11.
\end{equation}

For relativistic scalarons, $\zeta=\sqrt{{3}/{2}}$. Using this value in equation \eqref{DH3}, we obtain,
\begin{equation}\label{DH6}
	\left( \dfrac{\text{D}+{^3}\text{He}}{\text{H}}\right)^\text{Rel} \approx 5.27 \times 10^{-5}.
\end{equation}

The value of $\zeta$ for non-relativistic scalaron is calculated by considering the lower and upper bound of scalaron mass obtained through the BBN process. Using equation \eqref{BBN22}, we get, $\zeta_\text{Lower} \approx1$ (for $m_\psi^{\text{Lower}}$) and $\zeta_\text{Upper} \approx1.00004$ (for $m_\psi^{\text{Upper}}$). Using the above values of $\zeta$ in \eqref{DH3}, we obtain D+$^3$He abundance for non-relativistic scalarons (for both upper and lower bounds on scalaron mass) as,
\begin{equation}\label{DH9}
		\left( \dfrac{\text{D}+{^3}\text{He}}{\text{H}}\right)^\text{NRel} \approx 3.97 \times 10^{-5}. 
\end{equation}

The pre-solar abundance of $\left({\text{D}+{^3}\text{He}}\right)$ relative to hydrogen has been determined by the studies on solar wind abundance of $^3$He. In the last 4.5 Gyrs of history of the solar system, $^3$He has been deposited by solar wind on the moon, meteorites and breccias. Any pre-solar D present during the entry of the Sun on the main sequence has been burned into $^3$He. Therefore, observation of solar wind abundance of $^3$He is an indicator of the pre-solar abundance D+$^3$He \citep{1985ARA&A..23..319B}. This has been found as \citep{Olive:1999qe},
\begin{equation}\label{DH10}
	\left( \dfrac{\text{D}+{^3}\text{He}}{\text{H}}\right)_{\odot}^\text{pre-solar}=(4.1\pm0.6\pm1.4) \times 10^{-5}.
\end{equation}

Comparing the calculated values of $({\text{D}+{^3}\text{He}})/{\text{H}}$ for relativistic and non-relativistic scalarons (see equations \eqref{DH6} and \eqref{DH9}) with the observed pre-solar abundance (see equation \eqref{DH10}), we have seen that the predicted abundance of D+$^3$He in $f(R)$ gravity is compatible with the observed pre-solar bound.

\section{Results and Discussions}\label{sec4}

In this work $f(R)$ gravity scalaron is taken as additional density component in the Friedmann cosmology. Substituting modified expansion rate in the BBN epoch, we investigate the effect of scalarons on time-temperature relation and hence the elemental abundance. This allows us to produce bound on mass of the scalarons and PBHs.

We demonstrate that vacuum field equations in $f(R)$ gravity give rise to black hole solutions in cosmological background (Schwarzschild-de Sitter like). By extending the relation between scalaron mass and black hole mass obtained earlier through the consideration of curvature corrections to quantum vacuum fluctuations in black hole spacetime, we produce bound on mass range of PBHs. The usually expected PBH mass range ($10^{-38}-10^6\text{ }M_{\odot}$) gives rise to the scalaron mass in the range $10^{28}-10^{-16}$ eV (see equations \eqref{PBH11} and figure~\ref{fig1} (\emph{Left})). In order to investigate possibility of constraining scalaron mass through the BBN process we have tested unscreening of the scalarons over the scale of particle horizon at $t=1$ s. It has been found that all the scalarons in the above mass range are unscreened (i.e. $\psi_0/\phi>1$) across the particle horizon. The \emph{LPTK-Psi-Phi} plot generated for the considered PBH mass range is shown in figure~\ref{fig1} (\emph{Right}). The unscreened scalarons are taken as an additional source term in the expansion rate of the radiation dominated universe. The observed upper bound on the shift of freezeout temperature and elemental abundance of $^4$He have been employed to constrain the scalaron mass range. The range of scalaron mass which does not spoil the BBN era is found to be $10^{-16}-10^4\text{ eV}$ {for both relativistic and non-relativistic scalarons}. The corresponding PBH mass range is $10^{6}-10^{-14}\text{ }M_{\odot}$. Therefore, PBHs with $M<10^{-14}\text{ }M_{\odot}$ are found to be incompatible with BBN { study conducted in this work}. The mass window $10^3-10^6\text{ }M_{\odot}$ obtained in this work is eligible to provide with massive seeds required for the formation of supermassive black hole in the high redshift universe. The constrained range of PBH mass and scalaron mass is depicted in figure~\ref{fig2} (\emph{Left}). Figure~\ref{fig2} (\emph{Right}) displays the \emph{LPTK-Psi-Phi} plot for constrained PBH masses. PBHs with $M<10^5 M_{\odot}$ formed before the BBN epoch are often considered as non-baryonic in nature. Previous literature have discussed that mass windows for PBHs $10^{-13}-10^{-9}\text{ } M_{\odot}$ and $1-10^3\text{ }M_{\odot}$ provide with the potential candidates for dark matter density. In this study, we report the possibility of high mass PBHs with $M \approx 10^6 \text{ }M_{\odot}$. Therefore, the constrained mass range for PBHs reported in this work is eligible to provide PBH dark matter. On the other hand, if scalar field is an alternative to dark matter candidates (WIMPs) then $f(R)$ gravity scalarons which are consistent with BBN can provide with interesting dark matter candidates{, for} instance, axion like scalar fields are of potential interest. This study shows that the scalarons associated with PBHs of mass range $10^{-5}-10^{-10}\text{ }M_{\odot}$ are eligible for axion like particles $({10}^{-5}\text{ eV}\le m_\psi\le 1$ eV) and hence may qualify as a potential scalar field dark matter candidate.

The effect of $f(R)$ gravity scalarons on neutron and deuterium fraction is also investigated. The freezeout temperature for non-relativistic scalarons is found to be same as that of GR. However, it is increased by 8.3\% for relativistic scalarons (see equations \eqref{BBN28}). The deviation of neutron and deuterium fraction is, therefore, small ensuring GR like expansion history during the BBN epoch. GR like variation of neutron and deuterium fraction in scalaron gravity is shown in figures~\ref{fig3}.

Relative abundance of (D+$^3$He) is obtained in presence of scalarons. It is estimated for both relativistic and non-relativistic scalaron by taking present value of baryon-to-photon ratio as,  $\eta \approx 6.11 \times 10^{-10}$. For relativistic scalarons, (D+$^3$He) abundance is found as $ 5.27 \times 10^{-5}$, {whereas,} for non-relativistic scalarons, (D+$^3$He)/H $\approx 4\times 10^{-5}$. These abundances are found to be consistent with the observed pre-solar bound.

{
	
The range of PBH mass ($10^{-14}-10^6 \text{ }M_\odot$) obtained in this study is important for future astrophysical measurements in relation to gravitational waves (GWs). GW observations from LIGO-Virgo will provide with an opportunity to detect PBHs \citep{Carr:2023tpt, Escriva:2022bwe}. Four candidate black hole binaries are likely to be of sub-solar masses \citep{Phukon:2021cus}. If confirmed in more measurements, this will clearly indicate that these objects are not products of stellar evolution and hence may be of primordial origin. Moreover, the LIGO-Virgo confirmed black hole candidates with masses of  a few $100 \text{ }M_\odot$. These are also thought to be of primordial origin and not to be products of stellar core collapse. Very low mass compact astrophysical objects $(10^{-4}  -10^{-10} \text{ }M_\odot)$ observed through gravitational microlensing in the OGLE survey \citep{PhysRevD.99.083503} may be PBHs of the BBN window obtained here.   

BBN, however, cannot constrain ultralight PBHs, say those with $M \le 10^{-24} \text{ }M_\odot$. These ultralight PBHs are capable of causing small scale fluctuations in density before reheating the universe with Hawking radiation \citep{Papanikolaou:2020qtd}. These density fluctuations may leave signature through production of GW { and hence,} the observation of these GWs may constrain such scenarios containing ultralight PBHs. It has also been reported that GW spectrum of $10^{-25}  -10^{-29} \text{ }M_\odot$ PBHs are in the window accessible to LIGO and DECIGO \cite{Domenech:2020ssp}. $f(R)$ gravity has been found to affect the gravitational potential of a Poisson distributed assembly of PBHs which produces scalar induced GW. It is reported \citep{Papanikolaou:2021uhe} scalar induced GW can be used as a probe to constrain alternative gravitational theories.

}

The observed bound on freezeout temperature and abundances of primordial elements leaves room for testing interesting modification to GR. The small shift of the speedup factor ($\zeta$) of the cosmic expansion rate in scalaron gravity is indication of an early cosmic history which mimics the one expected in GR. We wish to infer that the PBH mass range derived from BBN in presence of scalarons is eligible to act as the non-baryonic dark matter component which is otherwise contemplated as made of unknown particle species within GR. The higher side of the PBH mass range is a precursor of high redshift supermassive black holes. Scalaron masses obtained through BBN consideration provide with an independent test of $f(R)$ gravity theory in a strong field environment ($\phi \sim 0.2$). The mass range accommodates the higher side ($10^{-16}-10^{-14}$ eV) of the one obtained through solar system constraint (see Fig. \ref{fig3}) on the PPN parameter $\gamma$. Therefore, BBN can be used as an independent strong field probe of gravitation theory which is unconstrained in local environment.

\acknowledgments

Corresponding author acknowledges Debojit Paul of Department of Physics, Gauhati University for his helpful remark on the scalaron mass vs black hole mass plot during preparation of the manuscript. The author would also like to thank Hitendra Sarkar of Department of Physics, Gauhati University, P.C Lalremruati of Department of Physical Sciences, Indian Institutes of Science Education and Research (IISER) Kolkata and Mukul Bhattacharya of Department of Physics, Penn State University for assisting in formatting of the manuscript.

% Bibliography

%% [A] Recommended: using JHEP.bst file
\bibliographystyle{JHEP}
\bibliography{biblio.bib}

\providecommand{\href}[2]{#2}\begingroup\raggedright\begin{thebibliography}{100}

\bibitem{PhysRevLett.26.1344}
S.W.~Hawking, \emph{{Gravitational Radiation from Colliding Black Holes}}, \href{https://doi.org/10.1103/PhysRevLett.26.1344}{\emph{Phys. Rev. D} {\bfseries 26} (1971) 1344}.

\bibitem{1974MNRAS.168..399C}
B.J.~{Carr} and S.W.~{Hawking}, \emph{{Black holes in the early Universe}}, \href{https://doi.org/10.1093/mnras/168.2.399}{\emph{Mon. Not. Roy. Astron. Soc.} {\bfseries 168} (1974) 399}.

\bibitem{doi:10.1073/pnas.2211215119}
M.M.~Waldrop, \emph{Primordial black holes could hold the key to dark matter mysteries}, \href{https://doi.org/10.1073/pnas.2211215119}{\emph{Proc. Natl. Acad. Sci.} {\bfseries 119} (2022) e2211215119}.

\bibitem{HAWKING1974}
S.W.~Hawking, \emph{Black hole explosions?}, \href{https://doi.org/10.1038/248030a0}{\emph{Nature} {\bfseries 248} (1974) 30}.

\bibitem{Volonteri2021TheOO}
M.~Volonteri, M.~Habouzit and M.~Colpi, \emph{The origins of massive black holes}, {\emph{Nat. Rev. Phys.} {\bfseries 3} (2021) 732 }.

\bibitem{1981MNRAS.194..639C}
B.J.~{Carr}, \emph{{Pregalactic black hole accretion and the thermal history of the universe}}, \href{https://doi.org/10.1093/mnras/194.3.639}{\emph{Mon. Not. Roy. Astron. Soc.} {\bfseries 194} (1981) 639}.

\bibitem{1976ApJ...206....1P}
D.N.~{Page} and S.W.~{Hawking}, \emph{{Gamma rays from primordial black holes.}}, \href{https://doi.org/10.1086/154350}{\emph{Astrophys. J.} {\bfseries 206} (1976) 1}.

\bibitem{1976ApJ...206....8C}
B.J.~{Carr}, \emph{{Some cosmological consequences of primordial black-hole evaporations.}}, \href{https://doi.org/10.1086/154351}{\emph{Astrophys. J.} {\bfseries 206} (1976) 8}.

\bibitem{1996ApJ...459..487W}
E.L.~{Wright}, \emph{{On the Density of Primordial Black Holes in the Galactic Halo}}, \href{https://doi.org/10.1086/176910}{\emph{Astrophys. J.} {\bfseries 459} (1996) 487} [\href{https://arxiv.org/abs/astro-ph/9509074}{{\ttfamily astro-ph/9509074}}].

\bibitem{Lehoucq:2009ge}
R.~Lehoucq, M.~Casse, J.M.~Casandjian and I.~Grenier, \emph{{New constraints on the primordial black hole number density from Galactic gamma-ray astronomy}}, \href{https://doi.org/10.1051/0004-6361/200911961}{\emph{Astron. Astrophys.} {\bfseries 502} (2009) 37} [\href{https://arxiv.org/abs/0906.1648}{{\ttfamily 0906.1648}}].

\bibitem{Kiraly1981}
P.~Kiraly, J.~Szabelski, J.~Wdowczyk and A.W.~Wolfendale, \emph{Antiprotons in the cosmic radiation}, \href{https://doi.org/10.1038/293120a0}{\emph{Nature} {\bfseries 293} (1981) 120}.

\bibitem{1991ApJ...371..447M}
J.H.~{MacGibbon} and B.J.~{Carr}, \emph{{Cosmic Rays from Primordial Black Holes}}, \href{https://doi.org/10.1086/169909}{\emph{Astrophys. J.} {\bfseries 371} (1991) 447}.

\bibitem{1996MNRAS.283..626B}
A.A.~{Belyanin}, V.V.~{Kocharovsky} and V.V.~{Kocharovsky}, \emph{{Gamma-ray bursts from the final stage of primordial black hole evaporation}}, \href{https://doi.org/10.1093/mnras/283.2.626}{\emph{Mon. Not. Roy. Astron. Soc.} {\bfseries 283} (1996) 626}.

\bibitem{Rubin:2001yw}
S.G.~Rubin, A.S.~Sakharov and M.Y.~Khlopov, \emph{{The Formation of primary galactic nuclei during phase transitions in the early universe}}, \href{https://doi.org/10.1134/1.1385631}{\emph{J. Exp. Theor. Phys.} {\bfseries 91} (2001) 921} [\href{https://arxiv.org/abs/hep-ph/0106187}{{\ttfamily hep-ph/0106187}}].

\bibitem{Carr:2018rid}
B.~Carr and J.~Silk, \emph{{Primordial Black Holes as Generators of Cosmic Structures}}, \href{https://doi.org/10.1093/mnras/sty1204}{\emph{Mon. Not. Roy. Astron. Soc.} {\bfseries 478} (2018) 3756} [\href{https://arxiv.org/abs/1801.00672}{{\ttfamily 1801.00672}}].

\bibitem{2020ApJ...897L..14Y}
J.~{Yang}, F.~{Wang}, X.~{Fan}, J.F.~{Hennawi}, F.B.~{Davies}, M.~{Yue} et~al., \emph{{P{\={o}}niu{\={a}}'ena: A Luminous z = 7.5 Quasar Hosting a 1.5 Billion Solar Mass Black Hole}}, \href{https://doi.org/10.3847/2041-8213/ab9c26}{\emph{Astrophys. J. Lett.} {\bfseries 897} (2020) L14} [\href{https://arxiv.org/abs/2006.13452}{{\ttfamily 2006.13452}}].

\bibitem{2019ApJ...880...77O}
M.~{Onoue}, N.~{Kashikawa}, Y.~{Matsuoka}, N.~{Kato}, T.~{Izumi}, T.~{Nagao} et~al., \emph{{Subaru High-z Exploration of Low-luminosity Quasars (SHELLQs). VI. Black Hole Mass Measurements of Six Quasars at 6.1 {\ensuremath{\leq}} z {\ensuremath{\leq}} 6.7}}, \href{https://doi.org/10.3847/1538-4357/ab29e9}{\emph{Astrophys. J.} {\bfseries 880} (2019) 77} [\href{https://arxiv.org/abs/1904.07278}{{\ttfamily 1904.07278}}].

\bibitem{Feng:2020kxv}
W.-X.~Feng, H.-B.~Yu and Y.-M.~Zhong, \emph{{Seeding Supermassive Black Holes with Self-interacting Dark Matter: A Unified Scenario with Baryons}}, \href{https://doi.org/10.3847/2041-8213/ac04b0}{\emph{Astrophys. J. Lett.} {\bfseries 914} (2021) L26} [\href{https://arxiv.org/abs/2010.15132}{{\ttfamily 2010.15132}}].

\bibitem{Inayoshi:2019fun}
K.~Inayoshi, E.~Visbal and Z.~Haiman, \emph{{The Assembly of the First Massive Black Holes}}, \href{https://doi.org/10.1146/annurev-astro-120419-014455}{\emph{Annu. Rev. Astron. Astrophys.} {\bfseries 58} (2020) 27} [\href{https://arxiv.org/abs/1911.05791}{{\ttfamily 1911.05791}}].

\bibitem{Bromm:2002hb}
V.~Bromm and A.~Loeb, \emph{{Formation of the first supermassive black holes}}, \href{https://doi.org/10.1086/377529}{\emph{Astrophys. J.} {\bfseries 596} (2003) 34} [\href{https://arxiv.org/abs/astro-ph/0212400}{{\ttfamily astro-ph/0212400}}].

\bibitem{Begelman:2006db}
M.C.~Begelman, M.~Volonteri and M.J.~Rees, \emph{{Formation of supermassive black holes by direct collapse in pregalactic halos}}, \href{https://doi.org/10.1111/j.1365-2966.2006.10467.x}{\emph{Mon. Not. Roy. Astron. Soc.} {\bfseries 370} (2006) 289} [\href{https://arxiv.org/abs/astro-ph/0602363}{{\ttfamily astro-ph/0602363}}].

\bibitem{2010ApJ...716.1397F}
K.~{Freese}, C.~{Ilie}, D.~{Spolyar}, M.~{Valluri} and P.~{Bodenheimer}, \emph{{Supermassive Dark Stars: Detectable in JWST}}, \href{https://doi.org/10.1088/0004-637X/716/2/1397}{\emph{Astrophys. J.} {\bfseries 716} (2010) 1397} [\href{https://arxiv.org/abs/1002.2233}{{\ttfamily 1002.2233}}].

\bibitem{2003ApJ...594L..71A}
N.~{Afshordi}, P.~{McDonald} and D.N.~{Spergel}, \emph{{Primordial Black Holes as Dark Matter: The Power Spectrum and Evaporation of Early Structures}}, \href{https://doi.org/10.1086/378763}{\emph{Astrophys. J. Lett.} {\bfseries 594} (2003) L71} [\href{https://arxiv.org/abs/astro-ph/0302035}{{\ttfamily astro-ph/0302035}}].

\bibitem{Rogers:2023ezo}
K.K.~Rogers, R.~Hlo\v{z}ek, A.~Lagu\"e, M.M.~Ivanov, O.H.E.~Philcox, G.~Cabass et~al., \emph{{Ultra-light axions and the S $_{8}$ tension: joint constraints from the cosmic microwave background and galaxy clustering}}, \href{https://doi.org/10.1088/1475-7516/2023/06/023}{\emph{JCAP} {\bfseries 06} (2023) 023} [\href{https://arxiv.org/abs/2301.08361}{{\ttfamily 2301.08361}}].

\bibitem{PhysRevLett.85.1158}
W.~Hu, R.~Barkana and A.~Gruzinov, \emph{{Fuzzy Cold Dark Matter: The Wave Properties of Ultralight Particles}}, \href{https://doi.org/10.1103/PhysRevLett.85.1158}{\emph{Phys. Rev. Lett.} {\bfseries 85} (2000) 1158}.

\bibitem{Carr:2020xqk}
B.~Carr and F.~Kuhnel, \emph{{Primordial Black Holes as Dark Matter: Recent Developments}}, \href{https://doi.org/10.1146/annurev-nucl-050520-125911}{\emph{Ann. Rev. Nucl. Part. Sci.} {\bfseries 70} (2020) 355} [\href{https://arxiv.org/abs/2006.02838}{{\ttfamily 2006.02838}}].

\bibitem{Clesse:2016vqa}
S.~Clesse and J.~Garc\'\i{}a-Bellido, \emph{{The clustering of massive Primordial Black Holes as Dark Matter: measuring their mass distribution with Advanced LIGO}}, \href{https://doi.org/10.1016/j.dark.2016.10.002}{\emph{Phys. Dark Univ.} {\bfseries 15} (2017) 142} [\href{https://arxiv.org/abs/1603.05234}{{\ttfamily 1603.05234}}].

\bibitem{PhysRevD.94.083504}
B.~Carr, F.~Kuhnel and M.~Sandstad, \emph{Primordial black holes as dark matter}, \href{https://doi.org/10.1103/PhysRevD.94.083504}{\emph{Phys. Rev. D} {\bfseries 94} (2016) 083504}.

\bibitem{PhysRevD.102.103512}
C.~Keith, D.~Hooper, N.~Blinov and S.D.~McDermott, \emph{Constraints on primordial black holes from big bang nucleosynthesis revisited}, \href{https://doi.org/10.1103/PhysRevD.102.103512}{\emph{Phys. Rev. D} {\bfseries 102} (2020) 103512}.

\bibitem{Starobinsky:1980te}
A.A.~Starobinsky, \emph{{A New Type of Isotropic Cosmological Models Without Singularity}}, \href{https://doi.org/10.1016/0370-2693(80)90670-X}{\emph{Phys. Lett. B} {\bfseries 91} (1980) 99}.

\bibitem{Capozziello:2003gx}
S.~Capozziello, V.F.~Cardone, S.~Carloni and A.~Troisi, \emph{{Curvature quintessence matched with observational data}}, \href{https://doi.org/10.1142/S0218271803004407}{\emph{Int. J. Mod. Phys. D} {\bfseries 12} (2003) 1969} [\href{https://arxiv.org/abs/astro-ph/0307018}{{\ttfamily astro-ph/0307018}}].

\bibitem{Sotiriou:2008rp}
T.P.~Sotiriou and V.~Faraoni, \emph{{f(R) Theories Of Gravity}}, \href{https://doi.org/10.1103/RevModPhys.82.451}{\emph{Rev. Mod. Phys.} {\bfseries 82} (2010) 451} [\href{https://arxiv.org/abs/0805.1726}{{\ttfamily 0805.1726}}].

\bibitem{PhysRevD.93.043511}
M.~Kusakabe, S.~Koh, K.S.~Kim and M.-K.~Cheoun, \emph{{Constraints on modified Gauss-Bonnet gravity during big bang nucleosynthesis}}, \href{https://doi.org/10.1103/PhysRevD.93.043511}{\emph{Phys. Rev. D} {\bfseries 93} (2016) 043511}.

\bibitem{Anagnostopoulos2022gej}
F.K.~Anagnostopoulos, V.~Gakis, E.N.~Saridakis and S.~Basilakos, \emph{{New models and big bang nucleosynthesis constraints in f(Q) gravity}}, \href{https://doi.org/10.1140/epjc/s10052-023-11190-x}{\emph{Eur. Phys. J. C} {\bfseries 83} (2023) 58} [\href{https://arxiv.org/abs/2205.11445}{{\ttfamily 2205.11445}}].

\bibitem{Asimakis:2021yct}
P.~Asimakis, S.~Basilakos, N.E.~Mavromatos and E.N.~Saridakis, \emph{{Big bang nucleosynthesis constraints on higher-order modified gravities}}, \href{https://doi.org/10.1103/PhysRevD.105.084010}{\emph{Phys. Rev. D} {\bfseries 105} (2022) 084010} [\href{https://arxiv.org/abs/2112.10863}{{\ttfamily 2112.10863}}].

\bibitem{Sultan:2022aoa}
A.M.~Sultan and A.~Jawad, \emph{{Compatibility of big bang nucleosynthesis in some modified gravities}}, \href{https://doi.org/10.1140/epjc/s10052-022-10860-6}{\emph{Eur. Phys. J. C} {\bfseries 82} (2022) 905}.

\bibitem{2020ApJ...893...31K}
S.~{Kalita}, \emph{{The Galactic Center Black Hole, Sgr A*, as a Probe of New Gravitational Physics with the Scalaron Fifth Force}}, \href{https://doi.org/10.3847/1538-4357/ab7af7}{\emph{Astrophys. J.} {\bfseries 893} (2020) 31}.

\bibitem{2021ApJ...909..189K}
S.~{Kalita}, \emph{{Scalaron Gravity near Sagittarius A*: Investigation of Spin of the Black Hole and Observing Requirements}}, \href{https://doi.org/10.3847/1538-4357/abded5}{\emph{Astrophys. J.} {\bfseries 909} (2021) 189}.

\bibitem{Borka:2015vqa}
D.~Borka, S.~Capozziello, P.~Jovanovi\'c and V.~Borka~Jovanovi\'c, \emph{{Probing hybrid modified gravity by stellar motion around Galactic Center}}, \href{https://doi.org/10.1016/j.astropartphys.2016.03.002}{\emph{Astropart. Phys.} {\bfseries 79} (2016) 41} [\href{https://arxiv.org/abs/1504.07832}{{\ttfamily 1504.07832}}].

\bibitem{PhysRevD.104.L101502}
I.~De~Martino, R.~della Monica and M.~De~Laurentis, \emph{{$f(R)$ gravity after the detection of the orbital precession of the S2 star around the Galactic Center massive black hole}}, \href{https://doi.org/10.1103/PhysRevD.104.L101502}{\emph{Phys. Rev. D} {\bfseries 104} (2021) L101502}.

\bibitem{2022ApJ...925..126L}
P.C.~{Lalremruati} and S.~{Kalita}, \emph{{Is It Possible to See the Breaking Point of General Relativity near the Galactic Center Black Hole? Consideration of Scalaron and Higher-dimensional Gravity}}, \href{https://doi.org/10.3847/1538-4357/ac3af0}{\emph{Astrophys. J.} {\bfseries 925} (2022) 126}.

\bibitem{2023EPJC...83..120K}
S.~{Kalita} and P.~{Bhattacharjee}, \emph{{Constraining spacetime metrics within and outside general relativity through the Galactic Center black hole (SgrA*) shadow}}, \href{https://doi.org/10.1140/epjc/s10052-023-11226-2}{\emph{Eur. Phys. J. C} {\bfseries 83} (2023) 120}.

\bibitem{PhysRevD.92.044009}
M.~Cataneo, D.~Rapetti, F.~Schmidt, A.B.~Mantz, S.W.~Allen, D.E.~Applegate et~al., \emph{{New constraints on $f(R)$ gravity from clusters of galaxies}}, \href{https://doi.org/10.1103/PhysRevD.92.044009}{\emph{Phys. Rev. D} {\bfseries 92} (2015) 044009}.

\bibitem{PhysRevLett.117.051101}
X.~Liu, B.~Li, G.-B.~Zhao, M.-C.~Chiu, W.~Fang, C.~Pan et~al., \emph{{Constraining $f(R)$ Gravity Theory Using Weak Lensing Peak Statistics from the Canada-France-Hawaii-Telescope Lensing Survey}}, \href{https://doi.org/10.1103/PhysRevLett.117.051101}{\emph{Phys. Rev. Lett.} {\bfseries 117} (2016) 051101}.

\bibitem{2017A&A...598A.132H}
A.~{Hammami} and D.F.~{Mota}, \emph{{Probing modified gravity via the mass-temperature relation of galaxy clusters}}, \href{https://doi.org/10.1051/0004-6361/201629003}{\emph{Astron. Astrophys.} {\bfseries 598} (2017) A132} [\href{https://arxiv.org/abs/1603.08662}{{\ttfamily 1603.08662}}].

\bibitem{Li:2015rva}
B.~Li, J.-h.~He and L.~Gao, \emph{{Cluster gas fraction as a test of gravity}}, \href{https://doi.org/10.1093/mnras/stv2650}{\emph{Mon. Not. Roy. Astron. Soc.} {\bfseries 456} (2016) 146} [\href{https://arxiv.org/abs/1508.07366}{{\ttfamily 1508.07366}}].

\bibitem{2017MNRAS.467.1569A}
P.~{Arnalte-Mur}, W.A.~{Hellwing} and P.~{Norberg}, \emph{{Real- and redshift-space halo clustering in f(R) cosmologies}}, \href{https://doi.org/10.1093/mnras/stx196}{\emph{Mon. Not. Roy. Astron. Soc.} {\bfseries 467} (2017) 1569} [\href{https://arxiv.org/abs/1612.02355}{{\ttfamily 1612.02355}}].

\bibitem{1981ApJ...243....8F}
D.~{Falik} and R.~{Opher}, \emph{{Production of primordial helium and deuterium as a strong-field test of gravitation theory}}, \href{https://doi.org/10.1086/158560}{\emph{Astrophys. J.} {\bfseries 243} (1981) 8}.

\bibitem{1973GReGr...4..435R}
N.~{Rosen}, \emph{{A bi-metric theory of gravitation}}, \href{https://doi.org/10.1007/BF01215403}{\emph{Gen. Relativ. Gravit.} {\bfseries 4} (1973) 435}.

\bibitem{ROSEN1974455}
N.~{Rosen}, \emph{A theory of gravitation}, \href{https://doi.org/https://doi.org/10.1016/0003-4916(74)90311-X}{\emph{Ann. Phys.} {\bfseries 84} (1974) 455}.

\bibitem{SupernovaSearchTeam:1998fmf}
{\scshape Supernova Search Team} collaboration, \emph{{Observational evidence from supernovae for an accelerating universe and a cosmological constant}}, \href{https://doi.org/10.1086/300499}{\emph{Astron. J.} {\bfseries 116} (1998) 1009} [\href{https://arxiv.org/abs/astro-ph/9805201}{{\ttfamily astro-ph/9805201}}].

\bibitem{SupernovaCosmologyProject:1998vns}
{\scshape Supernova Cosmology Project} collaboration, \emph{{Measurements of $\Omega$ and $\Lambda$ from 42 high redshift supernovae}}, \href{https://doi.org/10.1086/307221}{\emph{Astrophys. J.} {\bfseries 517} (1999) 565} [\href{https://arxiv.org/abs/astro-ph/9812133}{{\ttfamily astro-ph/9812133}}].

\bibitem{2007ApJ...659...98R}
A.G.~{Riess}, L.-G.~{Strolger}, S.~{Casertano}, H.C.~{Ferguson}, B.~{Mobasher}, B.~{Gold} et~al., \emph{{New Hubble Space Telescope Discoveries of Type Ia Supernovae at $z \ge 1$: Narrowing Constraints on the Early Behavior of Dark Energy}}, \href{https://doi.org/10.1086/510378}{\emph{Astrophys. J.} {\bfseries 659} (2007) 98} [\href{https://arxiv.org/abs/astro-ph/0611572}{{\ttfamily astro-ph/0611572}}].

\bibitem{RevModPhys.61.1}
S.~Weinberg, \emph{The cosmological constant problem}, \href{https://doi.org/10.1103/RevModPhys.61.1}{\emph{Rev. Mod. Phys.} {\bfseries 61} (1989) 1}.

\bibitem{RevModPhys.75.559}
P.J.E.~Peebles and B.~Ratra, \emph{The cosmological constant and dark energy}, \href{https://doi.org/10.1103/RevModPhys.75.559}{\emph{Rev. Mod. Phys.} {\bfseries 75} (2003) 559}.

\bibitem{PhysRev.124.925}
C.~Brans and R.H.~Dicke, \emph{{Mach's Principle and a Relativistic Theory of Gravitation}}, \href{https://doi.org/10.1103/PhysRev.124.925}{\emph{Phys. Rev.} {\bfseries 124} (1961) 925}.

\bibitem{PhysRevD.23.347}
A.H.~Guth, \emph{Inflationary universe: A possible solution to the horizon and flatness problems}, \href{https://doi.org/10.1103/PhysRevD.23.347}{\emph{Phys. Rev. D} {\bfseries 23} (1981) 347}.

\bibitem{Linde:1983gd}
A.D.~Linde, \emph{{Chaotic Inflation}}, \href{https://doi.org/10.1016/0370-2693(83)90837-7}{\emph{Phys. Lett. B} {\bfseries 129} (1983) 177}.

\bibitem{PhysRevD.37.3406}
B.~Ratra and P.J.E.~Peebles, \emph{Cosmological consequences of a rolling homogeneous scalar field}, \href{https://doi.org/10.1103/PhysRevD.37.3406}{\emph{Phys. Rev. D} {\bfseries 37} (1988) 3406}.

\bibitem{PhysRevLett.80.1582}
R.R.~Caldwell, R.~Dave and P.J.~Steinhardt, \emph{{Cosmological Imprint of an Energy Component with General Equation of State}}, \href{https://doi.org/10.1103/PhysRevLett.80.1582}{\emph{Phys. Rev. Lett.} {\bfseries 80} (1998) 1582}.

\bibitem{PhysRevLett.81.3067}
S.M.~Carroll, \emph{{Quintessence and the Rest of the World: Suppressing Long-Range Interactions}}, \href{https://doi.org/10.1103/PhysRevLett.81.3067}{\emph{Phys. Rev. Lett.} {\bfseries 81} (1998) 3067}.

\bibitem{PhysRevLett.82.896}
I.~Zlatev, L.~Wang and P.J.~Steinhardt, \emph{{Quintessence, Cosmic Coincidence, and the Cosmological Constant}}, \href{https://doi.org/10.1103/PhysRevLett.82.896}{\emph{Phys. Rev. Lett.} {\bfseries 82} (1999) 896}.

\bibitem{Will2001}
C.M.~Will, \emph{{The Confrontation between General Relativity and Experiment}}, \href{https://doi.org/10.12942/lrr-2001-4}{\emph{Living Rev. Relativ.} {\bfseries 4} (2001) 4}.

\bibitem{1982ApJ...253..908T}
J.H.~{Taylor} and J.M.~{Weisberg}, \emph{{A new test of general relativity - Gravitational radiation and the binary pulsar PSR 1913+16}}, \href{https://doi.org/10.1086/159690}{\emph{Astrophys. J.} {\bfseries 253} (1982) 908}.

\bibitem{1998ApJ...505..352S}
I.H.~{Stairs}, Z.~{Arzoumanian}, F.~{Camilo}, A.G.~{Lyne}, D.J.~{Nice}, J.H.~{Taylor} et~al., \emph{{Measurement of Relativistic Orbital Decay in the PSR B1534+12 Binary System}}, \href{https://doi.org/10.1086/306151}{\emph{Astrophys. J.} {\bfseries 505} (1998) 352} [\href{https://arxiv.org/abs/astro-ph/9712296}{{\ttfamily astro-ph/9712296}}].

\bibitem{PhysRevX.6.041015}
{\scshape LIGO Scientific Collaboration and Virgo Collaboration} collaboration, \emph{{Binary Black Hole Mergers in the First Advanced LIGO Observing Run}}, \href{https://doi.org/10.1103/PhysRevX.6.041015}{\emph{Phys. Rev. X} {\bfseries 6} (2016) 041015}.

\bibitem{PhysRevLett.116.061102}
{\scshape LIGO Scientific Collaboration and Virgo Collaboration} collaboration, \emph{{Observation of Gravitational Waves from a Binary Black Hole Merger}}, \href{https://doi.org/10.1103/PhysRevLett.116.061102}{\emph{Phys. Rev. Lett.} {\bfseries 116} (2016) 061102}.

\bibitem{PhysRevLett.122.101102}
{\scshape GRAVITY Collaboration} collaboration, \emph{{Test of the Einstein Equivalence Principle near the Galactic Center Supermassive Black Hole}}, \href{https://doi.org/10.1103/PhysRevLett.122.101102}{\emph{Phys. Rev. Lett.} {\bfseries 122} (2019) 101102}.

\bibitem{2018A&A...615L..15G}
{GRAVITY Collaboration}, R.~{Abuter}, A.~{Amorim}, N.~{Anugu}, M.~{Baub{\"o}ck}, M.~{Benisty} et~al., \emph{{Detection of the gravitational redshift in the orbit of the star S2 near the Galactic centre massive black hole}}, \href{https://doi.org/10.1051/0004-6361/201833718}{\emph{Astron. Astrophys.} {\bfseries 615} (2018) L15} [\href{https://arxiv.org/abs/1807.09409}{{\ttfamily 1807.09409}}].

\bibitem{2020A&A...636L...5G}
{GRAVITY Collaboration}, R.~{Abuter}, A.~{Amorim}, M.~{Baub{\"o}ck}, J.P.~{Berger}, H.~{Bonnet} et~al., \emph{{Detection of the Schwarzschild precession in the orbit of the star S2 near the Galactic centre massive black hole}}, \href{https://doi.org/10.1051/0004-6361/202037813}{\emph{Astron. Astrophys.} {\bfseries 636} (2020) L5} [\href{https://arxiv.org/abs/2004.07187}{{\ttfamily 2004.07187}}].

\bibitem{Starobinsky_2007}
A.A.~Starobinsky, \emph{{Disappearing cosmological constant in f(R) gravity}}, \href{https://doi.org/10.1134/s0021364007150027}{\emph{J. Exp. Theor. Phys. Lett.} {\bfseries 86} (2007) 157}.

\bibitem{Capozziello:2002rd}
S.~Capozziello, \emph{{Curvature quintessence}}, \href{https://doi.org/10.1142/S0218271802002025}{\emph{Int. J. Mod. Phys. D} {\bfseries 11} (2002) 483} [\href{https://arxiv.org/abs/gr-qc/0201033}{{\ttfamily gr-qc/0201033}}].

\bibitem{PhysRevD.70.043528}
S.M.~Carroll, V.~Duvvuri, M.~Trodden and M.S.~Turner, \emph{Is cosmic speed-up due to new gravitational physics?}, \href{https://doi.org/10.1103/PhysRevD.70.043528}{\emph{Phys. Rev. D} {\bfseries 70} (2004) 043528}.

\bibitem{PhysRevD.75.127502}
I.~Sawicki and W.~Hu, \emph{{Stability of cosmological solutions in $f(R)$ models of gravity}}, \href{https://doi.org/10.1103/PhysRevD.75.127502}{\emph{Phys. Rev. D} {\bfseries 75} (2007) 127502}.

\bibitem{Nojiri2010wj}
S.~Nojiri and S.D.~Odintsov, \emph{{Unified cosmic history in modified gravity: from F(R) theory to Lorentz non-invariant models}}, \href{https://doi.org/10.1016/j.physrep.2011.04.001}{\emph{Phys. Rept.} {\bfseries 505} (2011) 59} [\href{https://arxiv.org/abs/1011.0544}{{\ttfamily 1011.0544}}].

\bibitem{Nojiri:2013zza}
S.~Nojiri and S.D.~Odintsov, \emph{{Accelerating cosmology in modified gravity: from convenient $F(R)$ or string-inspired theory to bimetric $F(R)$ gravity}}, \href{https://doi.org/10.1142/S0219887814600068}{\emph{Int. J. Geom. Meth. Mod. Phys.} {\bfseries 11} (2014) 1460006} [\href{https://arxiv.org/abs/1306.4426}{{\ttfamily 1306.4426}}].

\bibitem{PhysRevD.85.124004}
D.~Borka, P.~Jovanovi\ifmmode~\acute{c}\else \'{c}\fi{}, V.B.~Jovanovi\ifmmode~\acute{c}\else \'{c}\fi{} and A.F.~Zakharov, \emph{{Constraints on ${R}^{n}$ gravity from precession of orbits of S2-like stars}}, \href{https://doi.org/10.1103/PhysRevD.85.124004}{\emph{Phys. Rev. D} {\bfseries 85} (2012) 124004}.

\bibitem{2018IJMPD..2741009Z}
A.F.~{Zakharov}, \emph{{The black hole at the Galactic Center: Observations and models}}, \href{https://doi.org/10.1142/S0218271818410092}{\emph{Int. J. Mod. Phys. D} {\bfseries 27} (2018) 1841009} [\href{https://arxiv.org/abs/1801.09920}{{\ttfamily 1801.09920}}].

\bibitem{PhysRevD.97.104068}
M.~De~Laurentis, I.~De~Martino and R.~Lazkoz, \emph{{Analysis of the Yukawa gravitational potential in $f(R)$ gravity. II. Relativistic periastron advance}}, \href{https://doi.org/10.1103/PhysRevD.97.104068}{\emph{Phys. Rev. D} {\bfseries 97} (2018) 104068}.

\bibitem{2022JCAP...03..007D}
R.~{Della Monica} and I.~{de Martino}, \emph{{Unveiling the nature of SgrA* with the geodesic motion of S-stars}}, \href{https://doi.org/10.1088/1475-7516/2022/03/007}{\emph{JCAP} {\bfseries 2022} (2022) 007} [\href{https://arxiv.org/abs/2112.01888}{{\ttfamily 2112.01888}}].

\bibitem{2007MNRAS.375.1423C}
S.~{Capozziello}, V.F.~{Cardone} and A.~{Troisi}, \emph{{Low surface brightness galaxy rotation curves in the low energy limit of R$^{n}$ gravity: no need for dark matter?}}, \href{https://doi.org/10.1111/j.1365-2966.2007.11401.x}{\emph{Mon. Not. Roy. Astron. Soc.} {\bfseries 375} (2007) 1423} [\href{https://arxiv.org/abs/astro-ph/0603522}{{\ttfamily astro-ph/0603522}}].

\bibitem{10.1007/978-3-319-02063-1_1}
S.~Capozziello, \emph{{Recovering Flat Rotation Curves and Galactic Dynamics From f(R)-Gravity}},  in \emph{Accelerated Cosmic Expansion}, C.~Moreno~Gonz{\'a}lez, J.E.~Madriz~Aguilar and L.M.~Reyes~Barrera, eds., (Cham), pp.~3--17, Springer International Publishing, 2014.

\bibitem{amendola_tsujikawa_2010}
L.~Amendola and S.~Tsujikawa, \emph{{Dark Energy: Theory and Observations}}, Cambridge University Press, Cambridge (2010).

\bibitem{1972CMaPh..25..152H}
S.W.~{Hawking}, \emph{{Black holes in general relativity}}, \href{https://doi.org/10.1007/BF01877517}{\emph{Commun. Math. Phys.} {\bfseries 25} (1972) 152}.

\bibitem{PhysRevLett.108.081103}
T.P.~Sotiriou and V.~Faraoni, \emph{{Black Holes in Scalar-Tensor Gravity}}, \href{https://doi.org/10.1103/PhysRevLett.108.081103}{\emph{Phys. Rev. Lett.} {\bfseries 108} (2012) 081103}.

\bibitem{PhysRevLett.83.2699}
T.~Jacobson, \emph{{Primordial Black Hole Evolution in Tensor-Scalar Cosmology}}, \href{https://doi.org/10.1103/PhysRevLett.83.2699}{\emph{Phys. Rev. Lett.} {\bfseries 83} (1999) 2699}.

\bibitem{2015CQGra..32u4002S}
T.P.~{Sotiriou}, \emph{{Black holes and scalar fields}}, \href{https://doi.org/10.1088/0264-9381/32/21/214002}{\emph{Class. Quant. Grav.} {\bfseries 32} (2015) 214002} [\href{https://arxiv.org/abs/1505.00248}{{\ttfamily 1505.00248}}].

\bibitem{2015IJMPD..2442014H}
C.A.R.~{Herdeiro} and E.~{Radu}, \emph{{Asymptotically flat black holes with scalar hair: A review}}, \href{https://doi.org/10.1142/S0218271815420146}{\emph{Int. J. Mod. Phys. D} {\bfseries 24} (2015) 1542014} [\href{https://arxiv.org/abs/1504.08209}{{\ttfamily 1504.08209}}].

\bibitem{PhysRevD.100.063014}
K.~Clough, P.G.~Ferreira and M.~Lagos, \emph{{Growth of massive scalar hair around a Schwarzschild black hole}}, \href{https://doi.org/10.1103/PhysRevD.100.063014}{\emph{Phys. Rev. D} {\bfseries 100} (2019) 063014}.

\bibitem{2020JCAP...09..026K}
M.~{Khodadi}, A.~{Allahyari}, S.~{Vagnozzi} and D.F.~{Mota}, \emph{{Black holes with scalar hair in light of the Event Horizon Telescope}}, \href{https://doi.org/10.1088/1475-7516/2020/09/026}{\emph{JCAP} {\bfseries 2020} (2020) 026} [\href{https://arxiv.org/abs/2005.05992}{{\ttfamily 2005.05992}}].

\bibitem{2019JCAP...06..038H}
L.~{Hui}, D.~{Kabat}, X.~{Li}, L.~{Santoni} and S.S.C.~{Wong}, \emph{{Black hole hair from scalar dark matter}}, \href{https://doi.org/10.1088/1475-7516/2019/06/038}{\emph{JCAP} {\bfseries 2019} (2019) 038} [\href{https://arxiv.org/abs/1904.12803}{{\ttfamily 1904.12803}}].

\bibitem{Sakharov:1967pk}
A.D.~Sakharov, \emph{{Vacuum quantum fluctuations in curved space and the theory of gravitation}}, \href{https://doi.org/10.1070/PU1991v034n05ABEH002498}{\emph{Dokl. Akad. Nauk Ser. Fiz.} {\bfseries 177} (1967) 70}.

\bibitem{ruzmaikina1970quadratic}
T.~Ruzmaikina and A.~Ruzmaikin, \emph{{Quadratic corrections to the Lagrangian density of the gravitational field and the singularity}}, {\emph{Soviet Phys. J. Exp. Theor. Phys.} {\bfseries 30} (1970) 372}.

\bibitem{Kalita_2018}
S.~Kalita, \emph{{Gravitational Theories near the Galactic Center}}, \href{https://doi.org/10.3847/1538-4357/aaadbb}{\emph{Astrophys. J.} {\bfseries 855} (2018) 70}.

\bibitem{2018EPJC...78..676C}
S.~{Carneiro} and J.C.~{Fabris}, \emph{{Scalar field black holes}}, \href{https://doi.org/10.1140/epjc/s10052-018-6161-x}{\emph{Eur. Phys. J. C} {\bfseries 78} (2018) 676} [\href{https://arxiv.org/abs/1808.04423}{{\ttfamily 1808.04423}}].

\bibitem{PhysRevD.91.084011}
A.~Arvanitaki, M.~Baryakhtar and X.~Huang, \emph{{Discovering the QCD axion with black holes and gravitational waves}}, \href{https://doi.org/10.1103/PhysRevD.91.084011}{\emph{Phys. Rev. D} {\bfseries 91} (2015) 084011}.

\bibitem{2015PhR...568....1J}
A.~{Joyce}, B.~{Jain}, J.~{Khoury} and M.~{Trodden}, \emph{{Beyond the cosmological standard model}}, \href{https://doi.org/10.1016/j.physrep.2014.12.002}{\emph{Phys. Rep.} {\bfseries 568} (2015) 1} [\href{https://arxiv.org/abs/1407.0059}{{\ttfamily 1407.0059}}].

\bibitem{2019ARA&A..57..335F}
P.G.~{Ferreira}, \emph{{Cosmological Tests of Gravity}}, \href{https://doi.org/10.1146/annurev-astro-091918-104423}{\emph{Annu. Rev. Astron. Astrophys.} {\bfseries 57} (2019) 335} [\href{https://arxiv.org/abs/1902.10503}{{\ttfamily 1902.10503}}].

\bibitem{PhysRevD.76.064004}
W.~Hu and I.~Sawicki, \emph{{Models of $f(R)$ cosmic acceleration that evade solar system tests}}, \href{https://doi.org/10.1103/PhysRevD.76.064004}{\emph{Phys. Rev. D} {\bfseries 76} (2007) 064004}.

\bibitem{PhysRevD.69.044026}
J.~Khoury and A.~Weltman, \emph{Chameleon cosmology}, \href{https://doi.org/10.1103/PhysRevD.69.044026}{\emph{Phys. Rev. D} {\bfseries 69} (2004) 044026}.

\bibitem{PhysRevLett.93.171104}
J.~Khoury and A.~Weltman, \emph{{Chameleon Fields: Awaiting Surprises for Tests of Gravity in Space}}, \href{https://doi.org/10.1103/PhysRevLett.93.171104}{\emph{Phys. Rev. Lett.} {\bfseries 93} (2004) 171104}.

\bibitem{2022ApJ...941..183L}
P.C.~{Lalremruati} and S.~{Kalita}, \emph{{Effect of Dark Matter Distribution on Scalaron Gravity near the Galactic Center Black Hole and Its Prospects}}, \href{https://doi.org/10.3847/1538-4357/aca071}{\emph{Astrophys. J.} {\bfseries 941} (2022) 183}.

\bibitem{2023IJMPD..3250021P}
D.~{Paul}, S.~{Kalita} and A.~{Talukdar}, \emph{{Unscreening of f(R) gravity near the galactic center black hole: Testability through pericenter shift below S0-2{\textquoteright}s orbit}}, \href{https://doi.org/10.1142/S0218271823500219}{\emph{Int. J. Mod. Phys. D} {\bfseries 32} (2023) 2350021}.

\bibitem{Casas:1990fz}
J.A.~Casas, J.~Garcia-Bellido and M.~Quiros, \emph{{Nucleosynthesis bounds on Jordan-Brans-Dicke theories of gravity}}, \href{https://doi.org/10.1142/S0217732392000409}{\emph{Mod. Phys. Lett. A} {\bfseries 7} (1992) 447}.

\bibitem{Lambiase:2012fv}
G.~Lambiase, \emph{{Constraints on massive gravity theory from big bang nucleosynthesis}}, \href{https://doi.org/10.1088/1475-7516/2012/10/028}{\emph{JCAP} {\bfseries 10} (2012) 028} [\href{https://arxiv.org/abs/1208.5512}{{\ttfamily 1208.5512}}].

\bibitem{Capozziello:2017bxm}
S.~Capozziello, G.~Lambiase and E.N.~Saridakis, \emph{{Constraining f(T) teleparallel gravity by Big Bang Nucleosynthesis}}, \href{https://doi.org/10.1140/epjc/s10052-017-5143-8}{\emph{Eur. Phys. J. C} {\bfseries 77} (2017) 576} [\href{https://arxiv.org/abs/1702.07952}{{\ttfamily 1702.07952}}].

\bibitem{Asimakis:2022kfk}
P.~Asimakis, E.N.~Saridakis, S.~Basilakos and K.~Yesmakhanova, \emph{{Big Bang Nucleosynthesis Constraints on f (T, T$_{G}$) Gravity}}, \href{https://doi.org/10.3390/universe8090486}{\emph{Universe} {\bfseries 8} (2022) 486} [\href{https://arxiv.org/abs/2209.01595}{{\ttfamily 2209.01595}}].

\bibitem{Bhattacharjee:2021hwm}
S.~Bhattacharjee, \emph{{BBN constraints on f(Q,T) gravity}}, \href{https://doi.org/10.1142/S0217751X22500178}{\emph{Int. J. Mod. Phys. A} {\bfseries 37} (2022) 2250017} [\href{https://arxiv.org/abs/2102.12921}{{\ttfamily 2102.12921}}].

\bibitem{Ichikawa:2007fa}
K.~Ichikawa, \emph{{Cosmological Constraint on the Effective Number of Neutrino Species}},  in \emph{{19th Rencontres de Blois on Matter and Energy in the Universe: From Nucleosynthesis to Cosmology}}, 6, 2007 [\href{https://arxiv.org/abs/0706.3465}{{\ttfamily 0706.3465}}].

\bibitem{PhysRevD.97.023502}
M.~Kawasaki, K.~Kohri, T.~Moroi and Y.~Takaesu, \emph{Revisiting big-bang nucleosynthesis constraints on long-lived decaying particles}, \href{https://doi.org/10.1103/PhysRevD.97.023502}{\emph{Phys. Rev. D} {\bfseries 97} (2018) 023502}.

\bibitem{Yeh:2022heq}
T.-H.~Yeh, J.~Shelton, K.A.~Olive and B.D.~Fields, \emph{{Probing physics beyond the standard model: limits from BBN and the CMB independently and combined}}, \href{https://doi.org/10.1088/1475-7516/2022/10/046}{\emph{JCAP} {\bfseries 10} (2022) 046} [\href{https://arxiv.org/abs/2207.13133}{{\ttfamily 2207.13133}}].

\bibitem{PhysRevD.67.063501}
J.P.~Kneller and G.~Steigman, \emph{{Big bang nucleosynthesis and CMB constraints on dark energy}}, \href{https://doi.org/10.1103/PhysRevD.67.063501}{\emph{Phys. Rev. D} {\bfseries 67} (2003) 063501}.

\bibitem{PhysRevD.68.123512}
S.~Nojiri and S.D.~Odintsov, \emph{Modified gravity with negative and positive powers of curvature: Unification of inflation and cosmic acceleration}, \href{https://doi.org/10.1103/PhysRevD.68.123512}{\emph{Phys. Rev. D} {\bfseries 68} (2003) 123512}.

\bibitem{2007IJGMM..04..115N}
S.~{Nojiri} and S.D.~{Odintsov}, \emph{{Introduction to Modified Gravity and Gravitational Alternative for Dark Energy}}, \href{https://doi.org/10.1142/S0219887807001928}{\emph{Int. J. Geom. Meth. Mod. Phys.} {\bfseries 04} (2007) 115} [\href{https://arxiv.org/abs/hep-th/0601213}{{\ttfamily hep-th/0601213}}].

\bibitem{Nojiri:2008nt}
S.~Nojiri and S.D.~Odintsov, \emph{{Dark energy, inflation and dark matter from modified F(R) gravity}}, {\emph{TSPU Bulletin} {\bfseries N8(110)} (2011) 7} [\href{https://arxiv.org/abs/0807.0685}{{\ttfamily 0807.0685}}].

\bibitem{Cembranos:2008gj}
J.A.R.~Cembranos, \emph{{Dark Matter from $R^2$ Gravity}}, \href{https://doi.org/10.1103/PhysRevLett.102.141301}{\emph{Phys. Rev. Lett.} {\bfseries 102} (2009) 141301} [\href{https://arxiv.org/abs/0809.1653}{{\ttfamily 0809.1653}}].

\bibitem{Katsuragawa:2017wge}
T.~Katsuragawa and S.~Matsuzaki, \emph{{Cosmic History of Chameleonic Dark Matter in $F(R)$ Gravity}}, \href{https://doi.org/10.1103/PhysRevD.97.064037}{\emph{Phys. Rev. D} {\bfseries 97} (2018) 064037} [\href{https://arxiv.org/abs/1708.08702}{{\ttfamily 1708.08702}}].

\bibitem{Katsuragawa:2016yir}
T.~Katsuragawa and S.~Matsuzaki, \emph{{Dark matter in modified gravity?}}, \href{https://doi.org/10.1103/PhysRevD.95.044040}{\emph{Phys. Rev. D} {\bfseries 95} (2017) 044040} [\href{https://arxiv.org/abs/1610.01016}{{\ttfamily 1610.01016}}].

\bibitem{maartens_durrer_2010}
R.~Maartens and R.~Durrer, \emph{Dark energy and modified gravity},  in \emph{Dark Energy: Observational and Theoretical Approaches}, P.~Ruiz-Lapuente, ed., p.~48–91, Cambridge University Press (2010).

\bibitem{Weinberg:1972kfs}
S.~Weinberg, \emph{{Gravitation and Cosmology}: {Principles and Applications of the General Theory of Relativity}}, John Wiley and Sons, New York (1972).

\bibitem{PhysRevLett.79.1588}
D.F.~Torres, H.~Vucetich and A.~Plastino, \emph{{Early Universe Test of Nonextensive Statistics}}, \href{https://doi.org/10.1103/PhysRevLett.79.1588}{\emph{Phys. Rev. Lett.} {\bfseries 79} (1997) 1588}.

\bibitem{doi:10.1126/science.aan8895}
R.W.~Pattie, N.B.~Callahan, C.~Cude-Woods, E.R.~Adamek, L.J.~Broussard, S.M.~Clayton et~al., \emph{Measurement of the neutron lifetime using a magneto-gravitational trap and in situ detection}, \href{https://doi.org/10.1126/science.aan8895}{\emph{Science} {\bfseries 360} (2018) 627}.

\bibitem{Dienes:2022zgd}
K.R.~Dienes, L.~Heurtier, F.~Huang, D.~Kim, T.M.P.~Tait and B.~Thomas, \emph{{Primordial Black Holes Place the Universe in Stasis}},  \href{https://arxiv.org/abs/2212.01369}{{\ttfamily 2212.01369}}.

\bibitem{2019JHEP...08..001H}
D.~{Hooper}, G.~{Krnjaic} and S.D.~{McDermott}, \emph{{Dark radiation and superheavy dark matter from black hole domination}}, \href{https://doi.org/10.1007/JHEP08(2019)001}{\emph{J. High Energy Phys.} {\bfseries 2019} (2019) 1} [\href{https://arxiv.org/abs/1905.01301}{{\ttfamily 1905.01301}}].

\bibitem{2020EPJP..135..552M}
I.~{Masina}, \emph{{Dark matter and dark radiation from evaporating primordial black holes}}, \href{https://doi.org/10.1140/epjp/s13360-020-00564-9}{\emph{Eur. Phys. J. Plus} {\bfseries 135} (2020) 552} [\href{https://arxiv.org/abs/2004.04740}{{\ttfamily 2004.04740}}].

\bibitem{2020JCAP...08..045B}
I.~{Baldes}, Q.~{Decant}, D.C.~{Hooper} and L.~{Lopez-Honorez}, \emph{{Non-cold dark matter from primordial black hole evaporation}}, \href{https://doi.org/10.1088/1475-7516/2020/08/045}{\emph{JCAP} {\bfseries 2020} (2020) 045} [\href{https://arxiv.org/abs/2004.14773}{{\ttfamily 2004.14773}}].

\bibitem{PhysRevD.43.984}
J.D.~Barrow, E.J.~Copeland, E.W.~Kolb and A.R.~Liddle, \emph{{Baryogenesis in extended inflation. II. Baryogenesis via primordial black holes}}, \href{https://doi.org/10.1103/PhysRevD.43.984}{\emph{Phys. Rev. D} {\bfseries 43} (1991) 984}.

\bibitem{2017PTEP.2017c3B02H}
Y.~{Hamada} and S.~{Iso}, \emph{{Baryon asymmetry from primordial black holes}}, \href{https://doi.org/10.1093/ptep/ptx011}{\emph{Prog. Theor. Exp. Phys.} {\bfseries 2017} (2017) 033B02} [\href{https://arxiv.org/abs/1610.02586}{{\ttfamily 1610.02586}}].

\bibitem{PhysRevD.103.043504}
D.~Hooper and G.~Krnjaic, \emph{{GUT baryogenesis with primordial black holes}}, \href{https://doi.org/10.1103/PhysRevD.103.043504}{\emph{Phys. Rev. D} {\bfseries 103} (2021) 043504}.

\bibitem{2003Natur.425..374B}
B.~{Bertotti}, L.~{Iess} and P.~{Tortora}, \emph{{A test of general relativity using radio links with the Cassini spacecraft}}, \href{https://doi.org/10.1038/nature01997}{\emph{Nature} {\bfseries 425} (2003) 374}.

\bibitem{Weinberg2008}
S.~Weinberg, \emph{Cosmology}, Oxford University Press Oxford, Oxford (2008).

\bibitem{1985ARA&A..23..319B}
A.M.~{Boesgaard} and G.~{Steigman}, \emph{{Big Bang nucleosynthesis: theories and observations.}}, \href{https://doi.org/10.1146/annurev.aa.23.090185.001535}{\emph{Annu. Rev. Astron. Astrophys.} {\bfseries 23} (1985) 319}.

\bibitem{Yu:2021djs}
H.~Yu, K.~Yang and J.~Li, \emph{{Constraints on running vacuum models with the baryon-to-photon ratio}}, \href{https://doi.org/10.1140/epjc/s10052-022-10164-9}{\emph{Eur. Phys. J. C} {\bfseries 82} (2022) 328} [\href{https://arxiv.org/abs/2103.02170}{{\ttfamily 2103.02170}}].

\bibitem{Olive:1999qe}
K.A.~Olive, \emph{{Primordial big bang nucleosynthesis}},  in \emph{{NATO Advanced Study Institute: Summer School on Theoretical and Observational Cosmology}}, 1, 1999 [\href{https://arxiv.org/abs/astro-ph/9901231}{{\ttfamily astro-ph/9901231}}].

\bibitem{Carr:2023tpt}
B.~Carr, S.~Clesse, J.~Garcia-Bellido, M.~Hawkins and F.~Kuhnel, \emph{{Observational Evidence for Primordial Black Holes: A Positivist Perspective}},  \href{https://arxiv.org/abs/2306.03903}{{\ttfamily 2306.03903}}.

\bibitem{Escriva:2022bwe}
A.~Escriv\`a, E.~Bagui and S.~Clesse, \emph{{Simulations of PBH formation at the QCD epoch and comparison with the GWTC-3 catalog}}, \href{https://doi.org/10.1088/1475-7516/2023/05/004}{\emph{JCAP} {\bfseries 05} (2023) 004} [\href{https://arxiv.org/abs/2209.06196}{{\ttfamily 2209.06196}}].

\bibitem{Phukon:2021cus}
K.S.~Phukon, G.~Baltus, S.~Caudill, S.~Clesse, A.~Depasse, M.~Fays et~al., \emph{{The hunt for sub-solar primordial black holes in low mass ratio binaries is open}},  \href{https://arxiv.org/abs/2105.11449}{{\ttfamily 2105.11449}}.

\bibitem{PhysRevD.99.083503}
H.~Niikura, M.~Takada, S.~Yokoyama, T.~Sumi and S.~Masaki, \emph{{Constraints on Earth-mass primordial black holes from OGLE 5-year microlensing events}}, \href{https://doi.org/10.1103/PhysRevD.99.083503}{\emph{Phys. Rev. D} {\bfseries 99} (2019) 083503}.

\bibitem{Papanikolaou:2020qtd}
T.~Papanikolaou, V.~Vennin and D.~Langlois, \emph{{Gravitational waves from a universe filled with primordial black holes}}, \href{https://doi.org/10.1088/1475-7516/2021/03/053}{\emph{JCAP} {\bfseries 03} (2021) 053} [\href{https://arxiv.org/abs/2010.11573}{{\ttfamily 2010.11573}}].

\bibitem{Domenech:2020ssp}
G.~Dom\`enech, C.~Lin and M.~Sasaki, \emph{{Gravitational wave constraints on the primordial black hole dominated early universe}}, \href{https://doi.org/10.1088/1475-7516/2021/11/E01}{\emph{JCAP} {\bfseries 04} (2021) 062} [\href{https://arxiv.org/abs/2012.08151}{{\ttfamily 2012.08151}}].

\bibitem{Papanikolaou:2021uhe}
T.~Papanikolaou, C.~Tzerefos, S.~Basilakos and E.N.~Saridakis, \emph{{Scalar induced gravitational waves from primordial black hole Poisson fluctuations in f(R) gravity}}, \href{https://doi.org/10.1088/1475-7516/2022/10/013}{\emph{JCAP} {\bfseries 10} (2022) 013} [\href{https://arxiv.org/abs/2112.15059}{{\ttfamily 2112.15059}}].

\end{thebibliography}\endgroup

%% or
%% [B] Manual formatting (see below)
%% (i) We suggest to always provide author, title and journal data or doi:
%% in short all the informations that clearly identify a document.
%% (ii) please avoid comments such as "For a review'', "For some examples",
%% "and references therein" or move them in the text. In general, please leave only references in the bibliography and move all
%% accessory text in footnotes.
%% (iii) Also, please have only one work for each \bibitem.

\end{document}